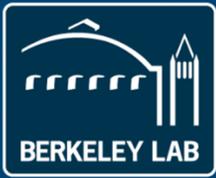

# BEAM
## Behavior Energy Autonomy Mobility

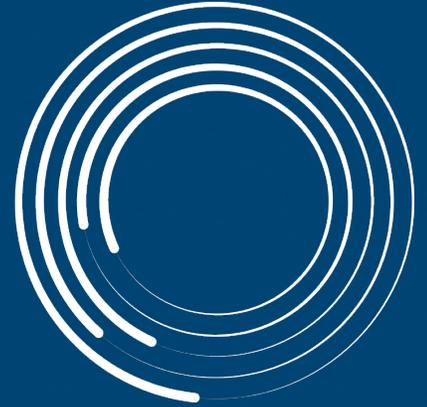

# BEAM: The Modeling Framework for Behavior, Energy, Autonomy & Mobility

The Open-Source Agent-Based Regional Transportation Model Unpacked: Concepts, Mechanisms, and Inner Dynamics

Haitam Laarabi, Zachary Needell, Rashid Waraich, Cristian Poliziani, Tom Wenzel

July 2023 — **Under Review**

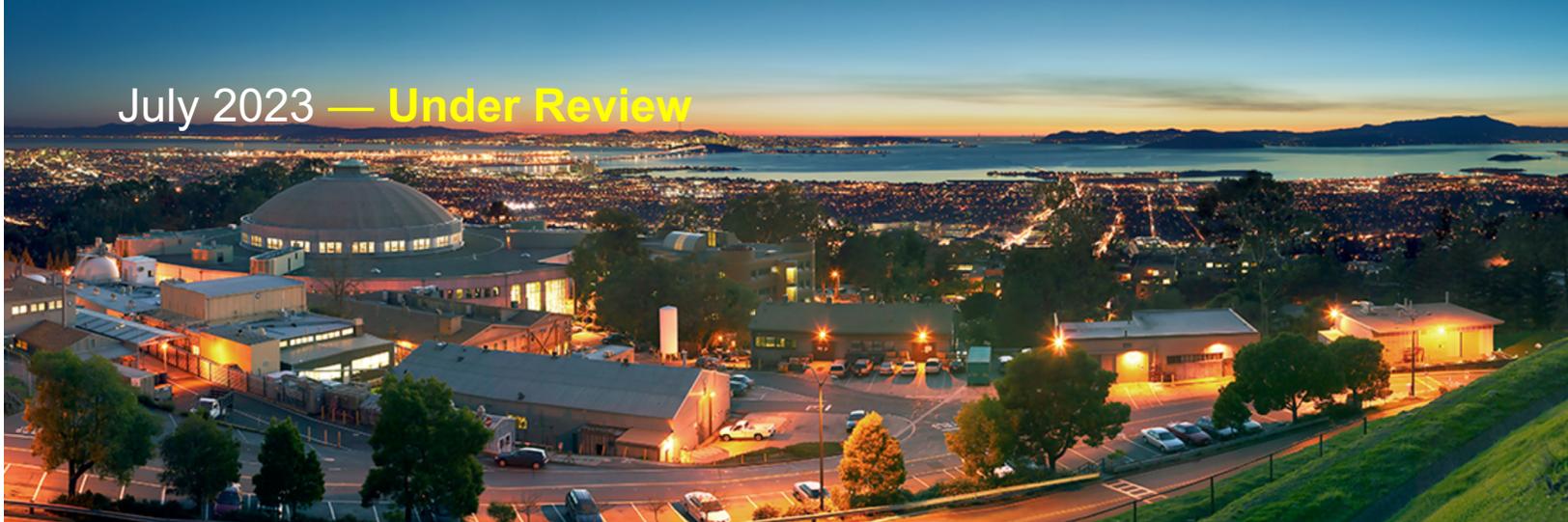

# BEAM:
# The Modeling Framework for Behavior, Energy, Autonomy & Mobility

Principal Authors
**Haitam Laarabi**
**Zachary Needell**
**Rashid Waraich**
**Cristian Poliziani**
**Tom Wenzel**

Lawrence Berkeley National Laboratory
1 Cyclotron Road, MS 90R4000
Berkeley CA 94720-8136

July 2023

# Acknowledgements

The work described in this study was conducted at Lawrence Berkeley National Laboratory and supported by the Vehicle Technologies Office (VTO) under the Systems and Modeling for Accelerated Research in Transportation (SMART) Mobility Laboratory Consortium, an initiative of the Energy Efficient Mobility Systems (EEMS) Program under Contract No. DE-AC02-05CH1123.

The authors would like to thank **Colin J.R. Sheppard**, **Michael Zilske**, **Sid Feygin**, **Xuan Jiang**, and all the collaborators from **Simrise** and the rest of the contributors (https://github.com/LBNL-UCB-STI/beam/graphs/contributors) for their contributions to the work on which this report is based.

**This report is under review!**

The authors thank the following experts for reviewing this report (affiliations do not imply that those organizations support or endorse this work):

[TBD: Reviewer Name]                                [Affiliation]

[TBD: Reviewer Name]                                [Affiliation]

[TBD: Reviewer Name]                                [Affiliation]

# Table of Contents

# Table of Figures

# List of Tables

# Acronyms and Abbreviations

| | |
|---|---|
| DOE | Department of Energy |
| ETA | Energy Technologies Area |
| EAEI | Energy Analysis and Environmental Impacts |
| STI | Sustainable Transportation Initiative |
| BEAM | The Modelling Framework for Behavior, Energy, Autonomy & Mobility |
| MATSim | The Multi-Agent Transport Simulation |
| BPR | Bureau of Public Roads |
| TAZ | Traffic Analysis Zones |
| CBG | Census Block Group |
| CBSA | Core Based Statistical Areas |
| INEXUS | Individual Experienced Utility-based Synthesis |
| NHTS | National Household Travel Survey |
| AMC | American Community Survey |
| PUMS | Public Use Micro |
| NYMTA | New York Metropolitan Transportation Authority |
| NJ Transit | New Jersey Transit |
| PATH | Port Authority of New York and New Jersey |
| EV | Electric Vehicle |
| CAV | Connected and Automated Vehicles |
| CACC | Coordinated Adaptive Cruise Control |
| OSM | Open Street Map |
| PBF | Protocolbuffer Binary Format |
| GTFS | General Transit Feed Specification |
| HPC | High Performance Computer |
| NERSC | National Energy Research Scientific Computing Center |
| AWS | Amazon Web Services |
| GCE | Google Cloud Engine |

# Executive Summary

This report provides a comprehensive description on the BEAM modeling framework, focusing on its design and structure, as well as a case study of its implementation in New York City.

**Conceptual Overview**

BEAM (Behavior, Energy, Autonomy, and Mobility) is a large-scale high-resolution transportation modeling framework that harnesses the principles of the actor model of computation to build a powerful and efficient agent-based model of travel behavior. It allows a detailed microscopic view of how people make travel choices and interact with the transportation system, enabling more accurate simulations of human mobility and urban transport networks. BEAM allows the analysis of numerous spatially defined but interacting layers (Figure ES-1), and integrates them into a cohesive representation of a regional transportation system. This integrated picture provides invaluable insights to policy makers and other stakeholders about how changes to the transportation system result in changes to traffic congestion, mode share, energy use, and emissions throughout a modeled region.

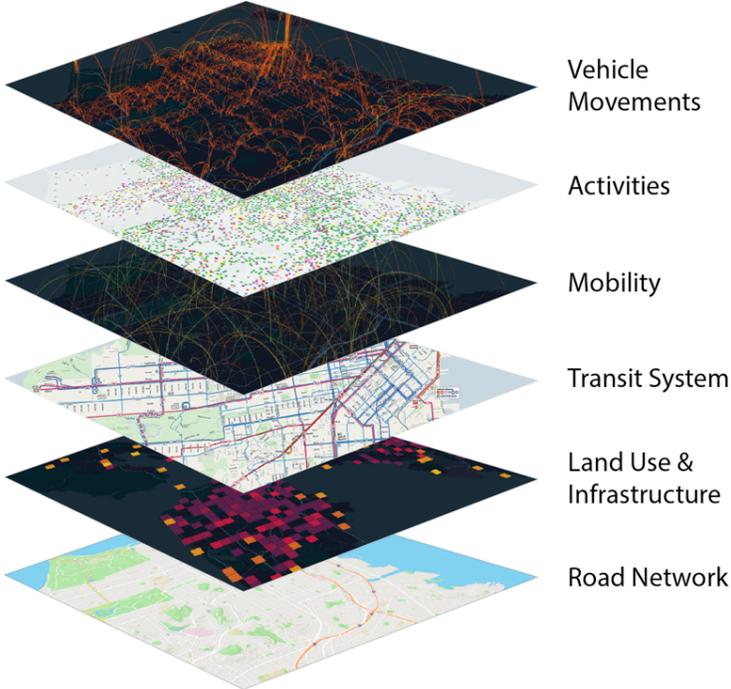

**Figure ES - 1. BEAM as a synthesis of multiple temporally and spatially resolved layers that together represent an integrated transportation system**

**Capabilities, Linkages, and Deployment**

BEAM is built around the open source MATSim codebase (Horni, Nagel, & Axhausen, 2016), with extensive modifications to allow for multithreaded within-day simulation of interacting agents. This report details the required and optional model inputs that can be used to define a scenario, including the road network, transit system, synthetic population, and input agent plans. Many of these inputs can be exported from existing MATSim models and read directly into BEAM, or they can be provided in simpler and more flexible formats. The report then describes the mechanics of BEAM's within-day AgentSim simulation, including features such as on-demand mobility, mode choice, parking selection, and discretionary trip planning. Finally, it summarizes BEAM's adaptation of MATSim's traffic network simulation and between-iteration replanning capabilities, the combination of which allow for BEAM to approximate dynamic user equilibrium across all of the choice dimensions offered its agents.

BEAM has always been designed as open-source software intended to benefit from an evolving ecosystem of related models and products. This report describes several key linkages to other, related models in the transportation and energy domain. These include activity-based travel demand models, which allow for a more sophisticated treatment of agents' pre-day travel planning, vehicle energy consumption models, metrics for a summarizing transport system's accessibility and efficiency, and integration with power grid models.

While BEAM is designed to flexibly and efficiently link with other models of transportation system components, it also includes the capability to build a BEAM implementation from scratch while relying only on free and publicly available data processed through several widely-used and open-source tools. This report describes these input data sources in detail and the steps required to turn them into a BEAM model that can be run locally or on cloud computing resources.

**Case Study**
Finally, a New York City case study is provided to showcase BEAM's application in a very large and intricate urban transportation system. Starting from scratch without relying on existing travel demand models, BEAM was employed to simulate the travel behavior of about 13 million inhabitants across five boroughs and nine counties. Utilizing tools like SynthPop and OpenStreetMap, the study carefully calibrated and validated the model to create a baseline scenario of typical weekday travel patterns in the New York City metro region. This model is responsive to real-world scenarios, such as adapting to the unprecedented changes in travel behavior and demands during various recovery phases of the COVID19 pandemic. The New York City study shows BEAM's ability to produce consistent and realistic outcomes given limited input data and fine-tuning, producing a flexible testbed for exploring major mobility shifts and policy impacts in complex urban environments.

**Conclusion**
BEAM's varied applications, flexibility, and integration capabilities make it a significant and valuable tool for various stakeholders in urban planning and transportation system analysis. Its unique ability to simulate individual behaviors, integrate with other models, and adapt to different real-world scenarios underscores its importance in the rapidly evolving field of transportation and emphasizes its potential as a valuable **proof-of-concept** tool to contribute to more informed and effective policy and planning decisions.

# 1. Introduction

## 1.1 Context

This report describes the BEAM model, an open-source agent-based transportation system model developed at Lawrence Berkeley National Laboratory. BEAM simulates the travel patterns of up to millions of individuals in a metropolitan area. The model simulates the travel of each agent across any travel mode available in a regional transportation system, including personal vehicles, public transit, new shared mobility services (such as ridehail, shared bikes and e-scooters), walking, and personal bikes. It simulates realistic interactions between agents and approximates a user equilibrium outcome, where the decisions made by each agent reflect both their own personal preferences and a realistic accounting of the impacts of the choices of all of the other agents on system outcomes. BEAM is designed to allow users and policymakers to understand the detailed operational and systemwide outcomes of different behavioral assumptions and scenarios in a richly detailed and realistic simulated transportation system. As a platform, BEAM is structured to balance detail, performance, and customizability in a behaviorally and operationally realistic manner. Despite the complexity of the integrated mobility systems being simulated, BEAM is designed with user-friendliness and accuracy in mind, allowing users to run meaningful simulations of tens of thousands of individual travelers (i.e., agents) on a personal computer, or deploy models of millions of agents in an HPC environment. BEAM offers a pipeline to allow users to build a model using publicly available data, and it also allows users to run scenarios defined by widely used open-source transportation modeling software.

In simulating an integrated multimodal transportation system, BEAM models a wide range of interacting systems as individual travelers progress through their day. It explicitly considers the daily travel patterns and preferences of individual travelers, and how their choices affect the performance of the road and transit networks that comprise a regional transportation system. In addition to conventional private vehicle travel, BEAM also simulates ride-hail fleet operations using either human-driven or centrally-managed automated vehicles (or both), personal or shared connected and automated vehicle (CAV) scheduling and use, and electric vehicle (EV) refueling requirements. BEAM incorporates these factors into a model of transportation network performance that can incorporate validated mesoscopic traffic network congestion modeling, the impacts of coordinated adaptive cruise control (CACC) and similar technologies on road network capacity, and traffic-dependent vehicle efficiency.

By simulating all of these interacting factors at once, BEAM approximates an equilibrium outcome that captures the complicated constraints associated with personal schedules and preferences, the transportation network and vehicle congestion on it, and operational realities of different modes, producing a picture that allows evaluation of the feasibility of different potential futures, and a better understanding of the directionality and relative strength of the relationship between different technology and policy developments and systemwide outcomes. By focusing on the travel of individual agents, BEAM also allows the analysis of the distributional impact of different scenarios on subgroups of the overall population, such as differentiating by housing location, demographic characteristics (such as income), and employment industry.

## 1.2 Other Models

For decades, researchers and planners have relied on the four-step modeling framework for simulating spatially and temporally resolved transportation networks (McNally, 2007). The four steps in this model—trip generation, trip distribution, mode choice, and (typically static) trip assignment—provide a robust and widely-used system for accounting for the expected impacts of small or gradual changes to the transport system such as population growth, highway expansion, or extension of a public transit system. The four-step model can be reasonably well suited to modeling changes that are focused on commute trips and do not deviate far from assuming privately owned cars as the primary mode of transport. Although it has proven less successful at capturing more nuanced behavioral factors and non-car-based modes. Over the past several decades, new types of transport models have been developed to address various shortcomings of the four-step model. On the demand side, the development of tour-based models and activity-based models (Rasouli & Timmermans, 2014) have brought more accuracy and responsiveness to the representation of individual and household travel decisions. On the supply side, advances such as dynamic traffic assignment (Peeta & Ziliaskopoulos, 2001) and car-following traffic modeling (Li & Sun, 2012) have greatly improved the accuracy of predicted road network travel times, and additional types of agent-based modeling has allowed for modelers to examine other modes, such as ride-hailing, with similar detail (Moniot, Borlaug, Ge, Wood, & Zimbler, 2022).

Several integrated models and modeling frameworks have emerged that capture both travel demand (in terms of trip, mode, and route choices) and network supply characteristics (in terms of car travel times and features of other modes, such as wait time, crowding, and number of transfers). In particular, the SimMobility Midterm (Lu, et al., 2015) and POLARIS (Auld, et al., 2016) models both offer a high-performance, agent-based representation of supply and demand that integrate a pre-day activity-based travel demand model with dynamic traffic assignment and traffic simulation, plus a within-day replanning module for unexpected discrepancies not accounted for in the pre-day choices. These models have been used to study the impacts of transit disruptions (Adnan, et al., 2017) and shared autonomous vehicles (Gurumurthy, de Souza, Enam, & Auld, 2020) (Nahmias-Biran, et al., 2019), and they offer a great deal of promise for future work. The integrated and closed structure of these models, often written in C-language, prioritizing performance over modularity, and using closed-source code, means that these models are typically not used (and almost never modified or adapted) without the participation of the initial model developers.

As a contrast, the MATSim modeling framework (W Axhausen, Horni, & Nagel, 2016) is designed to achieve maximum uptake via easy extensibility and modularity. MATSim, developed in Java, integrates easily with many open-source libraries and boasts an active developer community that regularly introduces new features, extensions, and plugins built around a core, open-source codebase. The core MATSim functionalities include modules to define and load a population of agents, containing their input "plans" that determine the activities and trips each agent intends to participate in over the course of a day. MATSim uses an iterative co-evolutionary algorithm to

BEAM: The Modeling Framework for Behavior, Energy, Autonomy & Mobility │2

approximate stochastic user equilibrium, where every iteration the agents seek to improve on their whole-day utility by either choosing a plan that has performed well for them in previous iterations or by following a new plan with different behaviors. Through several extensions, including mode and destination choice, the equilibrium plans in MATSim can approximate the constraints and feedbacks of activity-based modeling without relying on a formal, utility-based pre-day planning module (Horni, Scott, Balmer, & Axhausen, 2009) (Rakow & Nagel, 2023). MATSim also contains extensions that allow for the simulation of electric vehicles and on-demand mobility (Zwick & Axhausen, 2020) and electric vehicles (Waraich & Bischoff, 2016).

The BEAM model seeks to take advantage of many of the innovative aspects of MATSim, while making some structural changes that allow for BEAM to be better applied at scale to answer questions that can now only be answered using integrated travel demand models, and additional questions that cannot be answered at all using existing tools. In particular, BEAM focuses on improving the computational efficiency of MATSim, allowing simulations with more actors to be completed using fewer iterations and less run time. BEAM also directly incorporates within-day traveler choices and system responses into its behavioral simulations rather than accounting for these responses in between-iteration replanning, allowing for greater fidelity and faster convergence when modeling features, such as on-demand pooled mobility and shared vehicles, that require substantial interactions between the actors in the transportation system. Finally, BEAM is designed to ease direct integration with activity-based travel demand models, while still retaining the co-evolutionary structure that allows its agents to incorporate iteration-to-iteration learning that cannot be captured in such models.

In addition, BEAM is designed to allow the implementation of its open-source modeling code quickly in other metropolitan regions using only publicly available data, and allowing users superior flexibility in defining nuanced behavioral and policy scenarios, such as teleworking, capacity limitations on public transit, and the impacts of the COVID-19 pandemic on mode choice behavior.

## 1.3 BEAM Overview

Individual travelers, or "agents", in BEAM seek to maximize their experienced travel utility both by making decisions on-the-fly during a simulated travel day and by using an evolutionary algorithm to iteratively seek better transportation options over the course of several model iterations. This evolutionary algorithm (and much of the underlying Java codebase) is based on the one used in the popular MATSim software (Horni, Nagel, & Axhausen, 2016). In fact, BEAM has been developed on top of MATSim, while replacing most of its components with a new software architecture, discrete event simulation engine and operation models. After usually 10 to 20 iterations, BEAM reaches an approximate user equilibrium, where each agent's experienced travel outcomes roughly match her expectations, and where no agent can expect to significantly improve her outcomes by making a different choice. This equilibrium tends to be faster than in base MATSim because agents can have access to more real-time information when making their within-day choices, and (like in some MATSim extensions like in (Rakow & Nagel, 2023) agents can use a utility-maximizing approach while making choices rather than relying on uninformed or simplified choice structures.

As shown in Figure 1-1, BEAM is composed of several sub-components, three of which are considered the cornerstones of the framework: AgentSim, PhysSim, and the Replanning Module. AgentSim loads a synthetic population of households and individuals, from home/work locations in US Census data, and daily travel patterns derived from the National Household Travel Survey, and then models their behavior as they make their travel decisions at different times throughout the day and locations across the transportation network. In addition, AgentSim hosts the operation models of different ridehail service providers that matches drivers to trip requests in order to minimize wait time, as well as fleets of Autonomous Vehicles (AV). This capability can also be leveraged to model autonomous on-demand shuttles as first and last mile solutions for public transit (Poliziani et al., 2023). It also simulates micromobility and carsharing services that places and repositions shared bikes scooters and cars at locations to meet expected demand at different times of the day. AgentSim relies on a Router, typically the R5 routing engine (Pereira, Saraiva, Herszenhut, Braga, & Conway, 2021), or alternatively Graphhopper routing engine (Graphhopper, 2023), to produce a set of possible modes and routes for each traveler to choose among, based on the dynamic state of the transport network. PhysSim simulates travel times on the road network using the Java Discrete Event Queue Simulator (Waraich, Charypar, Balmer, & Axhausen, 2009) from MATSim. The Replanning Module revises the choices made by travelers from the previous day, offering them the opportunity to engage in a variety of optional activities, experiment with new modes of transport, or reiterate selections that proved successful in past iterations.

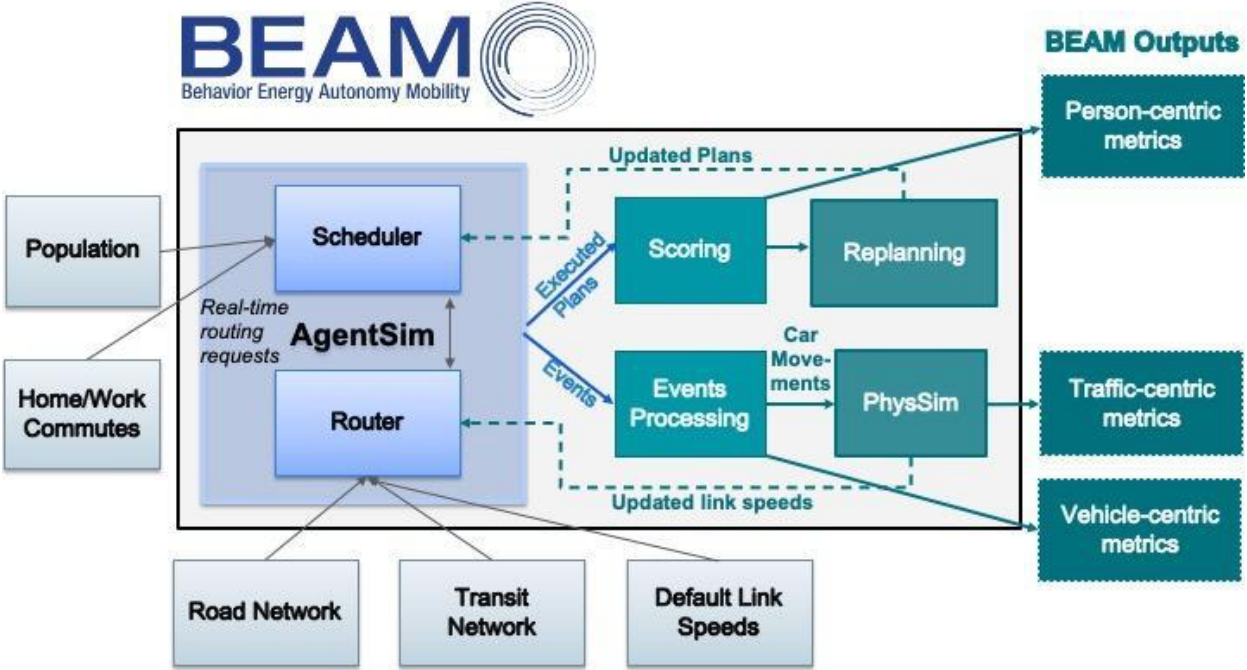

**Figure 1-1. Diagram of BEAM model structure.**

## 2. Model Structure

## 2.1 Actor System

BEAM relies on parallel asynchronous computation which is performed concurrently across CPU cores. BEAM is written primarily in the Scala language and uses the Akka library (Agha, 1986). Akka implements the Actor Model of Computation, which simplifies the process of utilizing high performance computing resources – in our case for deploying transportation simulations at full scale.

In BEAM, most aspects of the transportation system are defined as actors. These actors can represent individuals, such as Person Agents (travelers) and Ride Hail Agents (drivers of ride hail vehicles); they can represent resource managers that control aspects of the system, such as the Ride Hail Manager (that controls matching and repositioning of ride hail agents) and the Charging or Parking Manager (that matches vehicles with specific charging points or parking spots); and they can represent abstractions such as the Router (that provides agents with directions) and the scheduler (that centralizes the management of the simulation clock). These actors can operate independently, relying on no shared memory, and they communicate with each other through a tightly controlled message passing interface (Barker, 2015). Special trigger type messages have been defined, which are often sent by the scheduler to notify agents that it is time for them to perform a new action. An actor can also send a request message to another actor to ask for a specific set of information such as a route from origin to destination, which is provided through a response message.

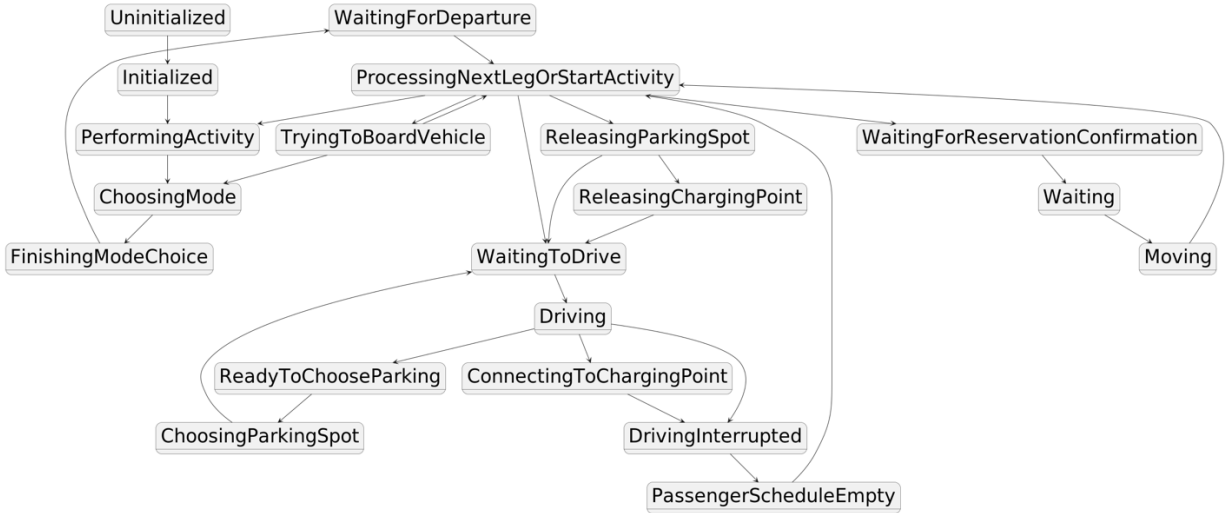

**Figure 2-1. Example of the possible states and state transitions accessible to the Person Agent.**

Some types of actors, specifically Person Agents and Ride Hail Agents, are BeamAgent actors. In the interest of conciseness, we will likewise refer to them as "agents". BeamAgent inherits the Akka FSM trait which provides a domain-specific language for programming agent actions as a finite state machine (Biermann & Feldman, 1972). These actors can exist only in a pre-specified set of states, and they transfer between states based on messages they receive, such as a request to pick up a passenger as seen in Figure 2-1. Because actors in BEAM can operate asynchronously, the rules for transitions between states and message handling must be robust to misaligned operations, such as two household members choosing the same vehicle at the same time. However, if misalignment occurs the agent logic in BEAM is designed to adapt to changing circumstances; for example, if an agent chose to use a bus but it is full; BEAM requires the agent to repeat the mode choice process and either wait for the next scheduled bus or choose a different mode.

It is important to note that BeamAgent actors do not have direct access to any internal clock. Instead, they rely on triggers passed to and from a BeamAgentScheduler actor to ensure a correct sequencing of events. For instance, if an agent begins participating in a planned activity that has an end time of 5:00 pm, it sends a message to the BeamAgentScheduler that schedules an ActivityEndTrigger that the BeamAgentScheduler will send back to the agent when the main simulation clock reaches 5:00 pm. The BeamAgent will eventually send a CompletionNotice back to the BeamAgentScheduler when it has completed all of its relevant actions in response to the trigger – in this case participating in mode choice and departing on the trip. The Scheduler advances the simulation clock in a rolling fashion – it maintains a time window of open triggers and only advances the time window forward when it has received CompletionNotices for all triggers in the portion of the time window being closed. When a sequence of events happens such that a trigger never gets completed, the BEAM simulation becomes stuck and cannot progress forward. This is a sign that some actor has entered an unplanned or unexpected state, a sign of a bug in the code that must be fixed. The version of BEAM released with this report has been tested thoroughly on large simulations and is thought to be less likely to contain such bugs, but the asynchronous nature of BEAM's actor system means that stuckness and other unexpected behaviors are common during the development and testing of new features.

What makes BeamAgents special is that they exhibit agency; i.e., they don't just change state but they have some degree of control or autonomy over themselves or other Agents. For example, a Ride Hail Manager can manage a fleet of actors as non-human drivers of autonomous vehicles. While travelers, fleet managers, or the Charging/Parking Network Managers are all Actors in BEAM, individual charging points, and parking stalls are treated as tools used by the BeamAgents, and are not considered agents or actors in BEAM. Vehicles can be either personal vehicles, bikes, shared cars, shared bikes/scooters, transit vehicles, or ride hail vehicles.

## 2.2 Model Inputs

### 2.2.1 Inputs Format

#### 2.2.1.1 Preferred Format

The typical format for importing various BEAM inputs such as population (households, persons, plans), vehicles (& vehicles types) and parking infrastructure is usually a CSV format. Except for the street network that relies preferably on Protocolbuffer Binary Format (PBF) and the General Transit Feed Specification (GTFS) for transit lines and schedules.

#### 2.2.1.2 Legacy MATSim Format

BEAM can also ingest input population data in the same xml format typically used by MATSim. MATSim population inputs require two xml files: a households file and a persons' file. The MATSim households file lists the income, location, and vehicles for each household (this format allows for a listing of the specific vehicle IDs for each household, rather than just the number of vehicles). The MATSim persons file contains information about both the person's attributes, such as age and sex, and their plans. Information on the storage and scoring of plans is given in Section 2.5 below.

### 2.2.2 Street Network

Spatial events in BEAM are associated with a street network. The street network in BEAM is represented by a graph where road links are modeled as edges and road intersections as graph nodes. Each street can be accessed by a specified set of modes (walk, car, bike, or bus). Furthermore, each link in the street network is associated with a capacity (its maximum throughput in terms of vehicles per time unit), a length, and a free flow speed.

The preferred method with which to load a street network into BEAM is through the Protocolbuffer Binary Format (PBF) typical to OpenStreetMap (OSM)[1]. OpenStreetMap networks can also be downloaded and simplified using the OSMnx tool[2], which improves the computational performance of routing algorithms by removing unnecessary nodes and links. For compatibility to MATSim, BEAM also offers utilities to convert a road network in the MATSIM ".xml" format to a compatible PBF file that can be read into BEAM. In addition to the road network, BEAM represents the rail and ferry network for public transit vehicles that do not traverse the road network but operate on dedicated right-of-way (e.g., subway, commuter rail, light rail, etc.). The combined transportation network of links, intersections, and transit stations are aggregated into Traffic Analysis Zones (TAZ); in many implementations of BEAM, TAZs are associated with census block groups, which are typically home to around 1,000 residents, spanning as few as several city blocks in dense urban locations and entire neighborhoods in less dense areas.

### 2.2.3 Transit Schedules

Transit operations in BEAM are simulated to match schedules that are input in the General Transit Feed Specification (GTFS) format. A GTFS archive contains a representation of every transit vehicle "trip" scheduled in a day, including the location, sequence, and timing of every stop along each trip, as well as information on fares and the locations of transfers. The R5 router in BEAM automatically synchronizes the OSM street network with the GTFS transit network, associating transit stations with walk access links and associating on-road mode legs such as bus with road links.

### 2.2.4 Population

Three types of population data, households, persons, and activity plans, are imported into BEAM to create a synthetic population that matches the aggregate socio-economic characteristics of the population in the modeling domain. The location, income, and number of vehicles for each household; the age, gender, and employment industry for each person; and the trips and activities that each person has planned for the travel day, are all fed into BEAM. Person and household ids are maintained so that activity plans can be assigned to each person, and persons can be aggregated into households. Each person's plans alternate between having a type "activity" and a type "leg." Activities have a type (such as "work" or "shopping"), are associated with specific locations, and have an end time. Trips do not need any additional information, but they can specify a trip mode, in which case the specified mode choice is treated as fixed in the first iteration of AgentSim.

---

[1] OpenStreetMap is an open-source mapping database that seeks to contain a representation of the entire worldwide road network, and various free tools exist to export OSM networks to PBF format for a given region. https://www.openstreetmap.org/

[2] OSMnx is a versatile Python package enabling users to easily download, model, project, visualize, and analyze a variety of geospatial data including urban networks and infrastructure from OpenStreetMap using minimal lines of code. https://osmnx.readthedocs.io/en/stable/

These input files can be generated using any census survey data such the National Household Travel Survey (NHTS) or from an existing implementation of ActivitySim (Galli, et al., 2009) in the area being studied. Their spatial resolution would depend on the granularity of the available data. Therefore, the resolution can be defined as an input to a BEAM simulation and can be at either the Census Block Group (CBG) or Tract level, a Traffic Analysis Zones (TAZ), H3 index (Uber, 2023) or customized spatial resolution.

### 2.2.5 Vehicles and Vehicle Types

Each vehicle simulated in BEAM must be associated with a given vehicle type, which determines various physical characteristics of that vehicle such as fuel type, energy storage capacity/range, passenger capacity, maximum speed, and energy consumption parameters – each of which is defined in a mandatory vehicle types input file. The type of individual vehicles can either be assigned probabilistically (with uniform probabilities, or probabilities that vary by household income), or the type of each vehicle can be enumerated directly through an input file called "vehicles.csv" that maps each vehicle to the corresponding vehicle type and initial state of charge. If the latter is not specified, BEAM will rely on a configuration parameter (meanPrivateVehicleStartingSOC) that indicates the mean for state of charge distribution.

**Table 2-1. Characteristics of the BEAM vehicle types input file**

| Characteristics | Description |
| --- | --- |
| Vehicle Type Id | Unique vehicle type ids |
| Seating Capacity, Standing Room Capacity, Length In Meter | Vehicle size characteristics |
| Primary/Secondary Fuel Type | Food (body vehicles), Gasoline, Diesel, Electricity, Biodiesel, Undefined |
| Primary/Secondary Fuel Consumption In Joule Per Meter | Average fuel consumption in joule per meter. It is used most of the time for estimation and can be used also for actual consumption. |
| Primary/Secondary Fuel Capacity In Joule | Tank or battery size, expressed in Joule |
| Primary/Secondary Vehicle Energy File | Relative path to a CSV file describing fuel consumption |
| Automation Level | Levels from 1 to 5. All vehicles that have a level of 3 and above will affect traffic flow using a CACC model. |

BEAM: The Modeling Framework for Behavior, Energy, Autonomy & Mobility │8

| Characteristics | Description |
| --- | --- |
| Vehicle Category | Body, Bike, Car, MediumDutyPassenger, LightDutyTruck, HeavyDutyTruck |
| Sample Probability String | ridehail \| all:P; income \| X-Y:Q |
| Charging Capability | label(Power\|Current) |

Vehicle Types input file, whose fields are detailed in Table 2-1, contains a vehicle type identifier and all relevant characteristics about the vehicle's size, energy consumption characteristics, and automation level. Optionally, users can also define an exhaustive list of all of the household vehicles to be simulated in BEAM in a "vehicles.csv" input file (Table 2-2), which is a useful way of defining the specific vehicle types owned by individual households as well as the initial state of charge distribution of the fleet. If this vehicles file is not included, vehicles are assigned to households with a probability that are defined in the configurations file. Transit vehicle types vehicles are assigned based on GTFS route type (e.g., bus, subway, rail) while agency-specific information about vehicle capacity and per mile fuel use can be input if available using the "TransitVehicleTypesByRoute.csv" file, which allows users to specify, for example, which routes use higher or lower capacity buses or trains, or assign vehicles with different fuel types and emissions characteristics to different routes.

**Table 2-2. Characteristics of the BEAM vehicle input file**

| Characteristics | Description |
| --- | --- |
| Vehicle Id | Unique vehicle ids |
| Vehicle Type Id | Vehicle type id from vehicle type files |
| Household Id | Id from household files |
| State Of Charge | Initial state of charge of the vehicle. It can be left empty for random assignment given the beam config parameter: beam.agentsim.agents.vehicles.meanPrivateVehicleStartingSOC beam.agentsim.agents.vehicles.meanRidehailVehicleStartingSOC |

### 2.2.6 On-Demand Fleets

On-demand fleets can either be defined explicitly or procedurally. When defining a fleet explicitly, the user uploads a .csv input containing one row for each on-demand vehicle. Each row contains a vehicle ID, a vehicle type, a starting location, a starting state of charge, and optionally a geofence beyond which the vehicle cannot operate. Currently, geofences can only be defined as circles with a predefined center and radius. Procedural input requires the user to specify a fleet size, defined as a ratio to the number of household vehicles defined in the input population – thus, entering a relative on-demand fleet size of 0.02 yields a scenario with one on-demand vehicle for every 50 household vehicles. These vehicles are initialized randomly into different TAZs with probabilities proportional to the number of home and work activities in each TAZ. The type of each vehicle is also chosen probabilistically, with probabilities as specified in the vehicleTypes.csv input file.

### 2.2.7 Parking and Charging Infrastructure

Parking and charging infrastructure are considered jointly in BEAM – each charging plug is associated with a single parking stall, but each parking stall does not necessarily have a plug. Parking infrastructure in BEAM is determined at the TAZ or road link level, where each TAZ (or link) can have any number of parking stalls with any set of characteristics. A parking stall's characteristics include its type (residential, workplace, or public), its pricing scheme (hourly or fixed), its cost (per hour or fixed), its charging plug type and power (if any), coordinates (optional) for link level resolution, the parking manager in charge of allocating its capacity in the format of MANAGER_TYPE(MANAGER_ID), and any time restrictions on its use in the format of VEHICLE_CATEGORY|HH:MM-HH:MM. When parking infrastructure is defined at the TAZ level, any number of parking stalls can be created in a TAZ with any set of characteristics. When a vehicle requests a stall within a TAZ, the parking manager creates a stall and assigns it to the user, with the distance between the requested destination and the parking stall's location inversely proportional to stalls availability. As a result, parking requests in TAZs with substantial parking availability will lead to very short walking legs between the stall and the destination, but in heavily-subscribed TAZs users may face long walks. When a vehicle leaves a stall, the stall is deleted and the unused parking availability in its TAZ is increased by one. When parking stalls are defined at the link level rather than TAZ level, parking stalls are created upon the initialization of the simulation (rather than upon request) and remain persistent throughout the simulation. These persistent stalls are allocated across road links within their TAZ, where they can be chosen, reserved, and released by agents over the course of the simulation.

## 2.3 AgentSim (Within-Day)

### 2.3.1 Activities, Trips and Mode Choice

The fundamental purpose of AgentSim is to model the outcomes when agents attempt to execute their stored daily travel plans, taking into consideration agents' origins and desired destinations; time constraints; access to different travel modes; the cost and duration, including access/egress and wait time, of different travel modes; and the constraints imposed on the transportation system of all the other agents in the simulation. This process is quite complex, as it contains much of the modeling work that is typically done in MATSim's replanning module, as well as many additional new features that have been developed from the ground up and not extended from MATSim, such as the parking choice and vehicle energy calculations. Two of the primary actors that mediate the actions of the different traveler agents in the simulation are the Scheduler and the Router, defined in more detail below.

#### 2.3.1.1 Scheduler

The Scheduler's purpose is to determine when each agent is scheduled to end an activity and then, when the simulation reaches that time, send a trigger to that agent telling it to begin its next trip. The scheduler also handles timing and triggers for other agents, such as transit driver agents and ride-hail driver agents.

#### 2.3.1.2 Router

The router calculates routes between locations on the travel network, for two purposes. First, to

BEAM: The Modeling Framework for Behavior, Energy, Autonomy & Mobility │ 10

generate travel times and costs using a skim lookup table which is a matrix of the duration, distance, and cost of travel between a pair of origin and destination TAZ (see Section 2.6.3). Agents use the skim table to estimate the utility of different modal options when choosing their preferred travel mode; the ride hail manager uses the skims table to estimate the time for available ridehail vehicles to reach travelers' ride requests). The skim table is also used to generate detailed routes for vehicles, which can then be fed into PhysSim to estimate traffic congestion levels on individual links and the overall road network given those trips.

In BEAM, person agents always travel in vehicles, and the trip between two activities can consist of legs associated with one or several unique vehicles. For the purposes of routing, a "body" is the vehicle type used for walk travel legs. For example, a personal vehicle trip can consist of an access walk leg from an activity location to the vehicle's parking location, a vehicle leg from the initial parking location to a parking location near the destination, and an egress walk leg from the parking location to the destination. Similarly, a transit trip includes either a walk, personal vehicle/bike, ridehail, or shared bike leg to access and egress a transit station/stop (if a household's personal vehicle is used to access a transit station, it is not available to egress the station at the end of the transit trip).

To build complete paths that consider multiple possible modes and a discrete set of potential access, trip, and egress vehicles, BEAM relies on the Conveyal R5 Routing Engine[3] an open-source tool implemented in Java. For every available mode, R5 searches over all available vehicles and routes to determine one or several optimal or nearly-optimal sequences of vehicle legs, as well as specific link-by-link routes for every leg, given a tradeoff between cost and travel time.

### 2.3.2 Mode Options

#### 2.3.2.1 Walk

Walking is a default mode available to all agents on all trips. Walkers are assumed to follow a configurable fixed speed and are routed along a network of walkable links. Elevation can be optionally considered when routing any type of vehicles including walking, by mapping the average gradient percent to each or some of the links in a CSV file indicated by the parameter "linkToGradePercentFilePath" in the BEAM configuration file.

#### 2.3.2.2 Personal Vehicle

When constructing a route alternative, R5 takes as input the set of personal vehicles accessible to the traveler, including any unused personal vehicles belonging to the traveler's household. R5 then calculates two routes for the nearest available car—a walking access leg from the agent's current location to the car, and a driving leg from the car's location to the trip's destination. In addition, the traveler sends a request to the parking manager to estimate the cost and walk access distance associated with parking at the trip's destination, given current parking availability. These components—the walk access distance, the driving time and toll cost, and the parking cost and egress distance—are all included in the agent's mode choice process. Because BEAM allocates personal vehicles to parking stalls at the beginning of the day and most households are often assumed to have dedicated parking stalls, BEAM assumes that most, but not all, walk legs from home to access a personal vehicle (either a light-duty vehicle or a personal bike) have a distance and duration of zero, when vehicles are parked at close enough distance to the home activity location. And if an agent takes a personal vehicle to a location, the return trip has to be made in that personal vehicle and not another mode.

---

[3] https://github.com/conveyal/r5

If the car mode is chosen, the vehicle is reserved for the agent and becomes unavailable for other travelers. The agent then begins a walk leg to the car's location, boards the car, and then begins a car trip to the destination. Once the agent reaches a configurable distance from the trip destination, the agent enters the parking choice process, sending out another query to the parking manager.

Upon entering the parking choice process, the agent sends another request to the parking manager, which returns a set of available parking stalls to the driver. BEAM includes several parking managers types, including one that assumes ubiquitous parking, another at link level where the parking stall is assigned, and a third that generates parking stall locations whose distance from the driver's destination is sampled based on the overall supply and demand of parking spaces at the TAZ level. This choice set of parking stalls can include stalls with different distances from the destination, different parking types (e.g., residential, workplace, or public), different prices, and different types of electric vehicle charging infrastructure. The agent chooses a parking stall through a configurable logit model that weighs monetary cost, walking access time, preference to home charging and charging requirements.

### 2.3.2.3 *Ride Hail*

BEAM can simulate one or several ridehail fleets of human-driven vehicles, similar to Lyft or Uber, or one or several centrally-managed fleets of autonomous vehicles, similar to Cruise or Waymo. A ride hail fleet gives travelers the option to request an on-demand vehicle to complete a trip. The user can request a solo trip, for which an empty vehicle will drive from its current location to the user's location, pick up the user, and take the user directly to a requested destination; or a user can request a pooled trip, where the matched vehicle could already have other riders, and any pooled user's trip can be diverted to pick up additional passengers. These solo and pooled options can be assigned different pricing models, and travelers can have configurable preferences and attitudes about them. To properly account for wait times, operational efficiency, energy consumption, and refueling/charging patterns, BEAM models the operation of these fleets in great detail. During AgentSim, BEAM models the shift timing and location of each vehicle, the matching between customers and vehicles, the queuing at the charging stations for autonomous ride-hail fleets, and the energy consumption and refueling behavior of on-demand vehicles.

If the traveler chooses the pooled shared ride option, a faster and sub-optimal version of the Alonso-Mora algorithm (Alonso-Mora, Samaranayake, Wallar, Frazzoli, & Rus, 2017) is used to pair that trip with other requests for pooled trips if possible (see Section 2.6.3 for more details). When they finish a trip in an area with low expected demand, ridehail vehicles follow one of various integrated repositioning heuristics to move to areas of higher anticipated demand. Assumptions regarding the timing and duration of ridehail driver shifts, and individual driver repositioning behavior between ride requests, can be refined to more accurately simulate ridehail driver behavior of a single, or even multiple, ridehail providers.

#### 2.3.2.4 Transit

When a traveler requests a transit route for a trip, BEAM's implementation of R5 uses a modified version (Pereira, Saraiva, Herszenhut, Braga, & Conway, 2021) of the McRAPTOR algorithm (Delling, Pajor, & Werneck, 2015) to quickly construct a set of possible Pareto optimal and slightly suboptimal routes on transit vehicles from origin to destination given a trip's departure time. The possible solutions are evaluated considering arrival time, total fare, and number of transfers. Each feasible transit route considers transit schedules defined by agency GTFS directories, ensuring that the returned itineraries are assigned to specific transit vehicle trips with feasible transfers. The cost of each possible transit itinerary is calculated considering a default set of fare rules that can be customized for an individual transit provider's unique system. In addition, BEAM can allow for non-transit vehicles to be used as access and/or egress legs for transit trips. By default, users consider any available personal vehicles or bicycles for access or egress, as well as ride-hail and (if enabled) shared vehicles, bikes, or scooters. Possible transit itineraries are grouped into a mode classification by the access and egress modes used: WALK_TRANSIT, BIKE_TRANSIT, DRIVE_TRANSIT and RIDE_HAIL_TRANSIT. Within these sets of itineraries, a multinomial logit model is used to choose the best itinerary for each mode classification, in approximation of a nested logit process. This transit route choice process allows for modelers to consider additional factors, such as preference for rail transit over bus, in transit routing even if they are not directly considered by the Router. The best itinerary for each mode classification is then returned to the requesting traveler for use in the mode choice process.

Individual transit vehicles and drivers are treated as BeamAgents, so that, at any given point on a vehicle's GTFS schedule between stops, the number of passenger agents on that vehicle can be calculated. If an agent tries to board a transit vehicle that is full of passengers, they will be denied access to that vehicle and have to replan their trip, either by waiting for the next transit vehicle or choosing a different mode. Currently BEAM uses a default maximum capacity for transit vehicles, by vehicle type (i.e., commuter or light rail, bus, etc.); however, BEAM can be configured to assign a different capacity by transit agency, vehicle type (e.g., number of subway, light rail, or commuter rail cars per train, or 40-foot vs 60-foot bus), route, or even run (e.g. commute vs. non-commute hours), based on detailed information provided by local transit agencies.

In BEAM, transit vehicles move between stations/stops based on the GTFS schedule. Bus schedules are not directly affected by traffic congestion on the road network, although the GTFS schedule could be revised to reflect expected bus arrival and departure times based on heavy congestion.

#### 2.3.2.5 CAV

Households in BEAM are modeled to maximize the utilization of any privately owned, as opposed to fleet operated, CAVs to meet their daily travel requirements. Each household follows a greedy heuristic to schedule the activity of any CAVs it owns in order to maximize the number of household members' trips that each CAV can cover. These household CAV schedules can involve empty trips between household members' activity locations, waiting for the next household member to depart from its activity to pick them up; when trips cannot be matched or congestion prevents a CAV from serving its scheduled trip, agents fall back on other modes to meet their travel needs.

### 2.3.3 Trip Mode Choice Model

BEAM gives users a rudimentary mode choice model that can be used by travelers in real time to choose the mode for upcoming trips based on the vehicles available to them and the real-time trip attributes returned to them by the router. Before agent $i$ departs for a trip, the utility of each available itinerary is calculated based on the cost $C_m$, duration $T_m$, and number of transfers $Transfer_m$ in that itinerary:

$$U_{m,i} = ASC_m + C_m + VOT_{m,i}T_m + \beta_{transfer}Transfer_m$$

where $ASC_m$ is the alternative-specific constant for mode $m$, $VOT_{m,i}$ is the agent's value of time for that trip, and $\beta_{transfer}$ is the utility parameter of a single transfer. A value of time can be defined for each agent in the scenario input; by default, BEAM infers the value of time for each mode based on the agent's household income. This agent-specific value of time can be configured to vary based on the mode of travel, and it can be tuned based on scenario requirements (for instance, it can vary by the level of crowding on a transit vehicle or the level of automation of a personal vehicle). Rather than taking the standard practice of fixing the standard deviation of the Gumbel distributed error term as one, instead we fix the cost coefficient at 1, setting the value of one unit of utility equal to one dollar and requiring that the magnitude of the error term, $\epsilon$, be specified as an input parameter. Thus, the probability of a user choosing alternative $m$ is:

$$P_{m,i} = \frac{e^{U_{m,i}/\epsilon}}{\sum_{n \in modes_i} e^{U_{n,i}/\epsilon}}$$

### 2.3.4 Tour Mode Choice Model

Optionally, BEAM also allows for travelers to participate in a tour mode choice process that considers the characteristics of all trips in an upcoming sequence of trips when determining the mode of the first trip in the sequence. In particular, this process lets the traveler choose between a BIKE_BASED or CAR_BASED tour (when a private vehicle is taken for the first trip and must be returned back to the tour's origin location at the end of the tour) or a WALK_BASED tour (where no private household vehicle is used but all other modal options are available for each trip). Tours by default are defined as sequences of trips that start and end when a traveler is at home, but trips in the input plans file can optionally be labeled with a "tour_id" identifier, allowing for more complicated (potentially nested) tour structures. Tour mode choice algorithm itself operates as a nested multinomial logit, where agents choose a tour mode based on the expected utility associated with choosing the best available trip mode for each trip, given the constraints set by the choice of tour mode:

$$U_{t,i} = \bar{\epsilon} \sum_{j \in Trips} \ln \left( \sum_{m \in modes_{t,i}} e^{U_{m,i,j}/\bar{\epsilon}} \right)$$

The expected maximum utility for agent $i$ choosing tour mode $t$ is defined as the sum of that tour mode's expected maximum utility for each trip $j$ on that tour. For a given trip, the expected maximum utility is the $logsum$ of the utility of that trip via all trip modes that will be available given tour mode $t$. For instance, a CAR_BASED tour would only have the CAR mode available, but a WALK_BASED tour would typically have WALK, RIDE_HAIL, and WALK_TRANSIT available. Note that if the traveler has access to a personal car at the beginning of the tour, the DRIVE_TRANSIT mode will be available for only the first and last trips in a WALK_BASED tour (similar for bikes and BIKE_TRANSIT), and if shared vehicles are available BIKE or CAR modes can be available for WALK_BASED tours as well. The tour mode is chosen based on a multinomial logit with respect to the tour mode utilities, and the choice of tour mode becomes a constraint on the available trip modes for each trip on the tour. To avoid overwhelming the router with routing requests, the utilities of each trip during tour mode choice are filled in based on the TAZ skims (see Section 2.6.3 below).

To preserve backwards compatibility, the tour mode choice capability is currently preserved in a separate branch of BEAM (with both branches receiving regular updates). In future releases of BEAM these two branches will be merged into a single branch with the ability to turn on or off tour mode choice.

### 2.3.5 Parking/Charging Choice Model

BeamAgents that are in a state requiring them to park a personal vehicle can request a parking stall from the Actor managing the parking infrastructure network. BeamAgents driving a plug-in battery (BEV) or hybrid (PHEV) electric vehicle send an inquiry to the Actor responsible for the parking and charging infrastructure. The received parking inquiry will go through the multinomial logit parking/charging choice model that samples different types of parking infrastructure and weighs each one of them based on the BeamAgent's predefined sensitivities towards parking cost, distance from destination, range anxiety (for electric vehicles), preference to residential parking, and detour distance for enroute charging. These sensitivities are input parameters to BEAM and can be tweaked to calibrate the electric load by charger type and the charging behavior against observed data.

### 2.3.6 Ride-hail Fleet Management

The central manager for the ridehail module in BEAM is the Actor RideHailManager, which is responsible for storing data and skims (more details about skims in Section 2.6.3), and keeping track of the state of the entire ridehail fleet; it creates the ridehail driver agents and their vehicles,

BEAM: The Modeling Framework for Behavior, Energy, Autonomy & Mobility | 15

and initializes ridehail vehicles remaining range (state of charge for EVs), initial locations, and driving shift timing and durations. It manages the routing requests and responses, the charging decision using the module DefaultRideHailDepotParkingManager, the repositioning strategies with the module RideHailRepositioning and the different customer matching algorithms with RideHailMatching. The RideHailManager also calculates the fares, handles customer inquiries and reservations, passenger vs empty (referred to as deadhead) VMT, idle vehicles, and stuck agents. Inquiries and reservations are two different types of customer requests for ridehail. An agent uses an inquiry to evaluate cost and wait/travel time for ridehail versus other modes, as a basis for their initial mode choice. After an agent has decided to use ridehail, they send a reservation request to be picked up by a ridehail vehicle. If a ride-hail passenger is unable to proceed to their next planned activity, it is considered "stuck" at the current activity and is removed from the simulation to avoid stalling the other agents. A large number of stuck agents is usually a sign of a bug in the computer code, a workflow problem, or unexpected inputs.

RideHailAgent, which is a BeamAgent representing a ridehail vehicle, can be of two different types: with a human driver, or an autonomous vehicle in the case of centrally-managed autonomous fleets. There are two main differences in the workflow between these two types of agents. Human drivers behave exactly as any Person Agent in the BEAM simulation; they are initially placed according to expected (or observed) demand; after completing a requested ride they either park or relocate to an area with expected future demand to await their next ride request; and they decide about when and where to refuel or charge their vehicles when necessary. At the end of their shift, they are removed from the simulation (that is, their commute to start or end their driving shift is not included). In contrast, autonomous vehicles can be dispatched, removed, and relocated between ride requests, based on expected or observed demand, and their charging decision is optimized by RideHailManager, which dispatches them to specific depots for refueling/charging. BEAM can simulate several independent ridehail service providers simultaneously, with human drivers vs autonomous vehicles, different pricing structures, etc. The input parameters for customizing each of the ridehail services (such as defaultCostPerMile, pooledCostPerMile and rideHailManager.radiusInMeters) can be found in the BEAM configuration file under beam.agentsim.agents.rideHail.

Currently the RideHailMatching treats solo trips and pooling trips as two separate processes. To pool customers, BEAM encompasses different matching algorithms that are based on AlonsoMora but have different tradeoffs in terms of runtime and level of optimality of the results. These matching algorithms can be calibrated using three main input parameters, which are "maxWaitingTimeInSec", "maxExcessRideTime" and "maxRequestsPerVehicle". The first two parameters put constraints on the maximum waiting time and the maximum acceptable detour time from a direct trip with a solo passenger. The last parameter speeds up the execution of the algorithm by reducing the solution space into a smaller chosen set of requests to match with a ridehail vehicle. Currently three ridehail matching algorithms can be utilized in BEAM:
- "MIP_ASSIGNMENT" which offers optimal vehicle-customer matching but the runtime is exponential, rendering it impractical for significantly large simulations.
- "ASYNC_GREEDY_ASSIGNMENT" which Is a greedy version of the AlonsoMora algorithm with parallelization of certain steps in the algorithm and represents the closest to the original algorithm i.e., "MIP_ASSIGNMENT".
- and "VEHICLE_CENTRIC_MATCHING" which is significantly different from AlonsoMora algorithm; while it maintains similarities in terms of how the graph of matching are created, it is the fastest pooling algorithm in BEAM, which parallelizes the process of every vehicle matching independently from others.

The assignment of ridehail vehicles to solo ridehail trip requests happens after all customers requesting a pooled ridehail trip have been matched; a rudimentary approach is used to match the closest available vehicle to the customer requesting a solo ridehail trip. This approach favors pooling over solo requests by first matching vehicles for pooling requests before matching solo requests. In addition to the matching, BEAM also comes with different reposition algorithms under RideHailRepositioning, which contains several different repositioning strategies the user can select and adjust their sensitivities using input parameters under "rideHail.repositioningManager", such:
- "DEFAULT_REPOSITIONING_MANAGER", which does not reposition the vehicles (that is, ridehail vehicles park after their last completed ride and await a subsequent ride request).
- "DEMAND_FOLLOWING_REPOSITIONING_MANAGER" which identifies the imbalance between simulated demand and supply, and redistributes idle vehicles to locations where ridehail demand is higher than supply.
- "REPOSITIONING_LOW_WAITING_TIMES" which relocates vehicles to locations where simulated waiting times are longer.
- and "INVERSE_SQUARE_DISTANCE_REPOSITIONING_FACTOR" which serves the same purpose as the demand following algorithm, but instead of relying on clustering, this strategy is sensitive to the inverse square law of the distance between simulated supply and demand in deciding where to relocate the vehicles. The pros of this algorithm is its faster runtime, and especially it reduces significantly the number parameters one needs to calibrate repositioning behavior.

### 2.3.7 Vehicle Sharing

The vehicle sharing module in BEAM is free floating, meaning that there are no fixed vehicle hubs; this can be used to simulate one-way shared vehicle service, such as GIG Carshare in the Bay Area, which allows vehicles to be parked in any legal parking space, or dockless bike/e-scooter sharing. Any limitations on where vehicles can be picked up or dropped off are handled through parking limitations; for instance, for a docked shared bike system, the only parking spots available to the bikes are at docking stations, or for a round-trip vehicle sharing service (such as GetAround or ZipCar), where vehicles must be returned to the location where they were picked up. The module is generic enough that it can handle all kinds of shared vehicles, including round-trip and one-way carsharing, and docked and dockless shared bikes and scooters, and can simulate multiple independent vehicles sharing services simultaneously. At initialization the vehicles are parked at stalls or at charging points depending on their power trains and availability of stalls. The module offers three types of vehicles sharing strategies that the user can select:
- fixed-non-reserving-fleet-by-TAZ: The initial locations of vehicles are either provided as input to BEAM or is distributed over the zoning level adopted by the simulation, such TAZs

- or CBGs.
- fixed-non-reserving: The initial locations of vehicles are distributed based on the distribution of in the home location of the population.
- inexhaustible-reserving: offers unlimited supply of vehicles, which can be used to estimate the maximum potential demand for vehicle sharing, as well as debug the vehicle sharing module in BEAM.

For docked bikeshare programs, bikes are initially located in docks at actual docking station locations, and can only be returned to a docking station that has an unoccupied dock. For dockless micromobility programs, bikes and scooters can be left as close to the agent's destination as possible.

## 2.4 Physsim (Traffic Network)

### 2.4.1 JDEQSim

BEAM utilizes the traffic network model from MATSim (called JDEQSim), an event-driven queue-based microsimulation of traffic flow where, instead of advancing in time steps, links are treated as finite-capacity queues. Vehicles are transmitted from one link to a subsequent link, and the transmission is treated as a discrete event, with the time between successive transmission events for a vehicle determined by the number of vehicles and capacity of each link. This framework allows for the efficient simulation of traffic flow on a large network while still capturing effects such as spillover and gridlock that are difficult to capture with simpler link-based methods (Waraich, Charypar, Balmer, & Axhausen, 2009).

### 2.4.2 Modifications to JDEQSim

Under different simulation scenarios, the roadway capacity of specific links can be increased, to simulate different portions of vehicles with cooperative adaptive cruise control (CACC, level 3 or above) on the road, or reduced, such as by removing a general travel lane and rededicating it to only transit vehicles. Since BEAM is based on a polynomial regression a la (Liu, Kan, Shladover, Lu, & Ferlis, 2018) interactions between human driven and autonomous CACC enabled vehicles can be simulated. Figure 2-2 shows a comparison between the original CACC model and simulation of CACC in BEAM.

BEAM also allows for developers to add other features that can impact the travel time on a link. For example, BEAM has a (currently unused) PickUpDropOffHolder class that counts the number of TNC vehicle boarding or alighting events on each link at different time periods and gives JDEQSim access to this value. Thus, with a user-defined empirical relationship, BEAM is capable of capturing the impact of TNC pickup/dropoff behavior on traffic congestion. This capability could be easily extended to account for the impacts of parking (or double parking), freight loading, or transit operations on vehicle flow as well.

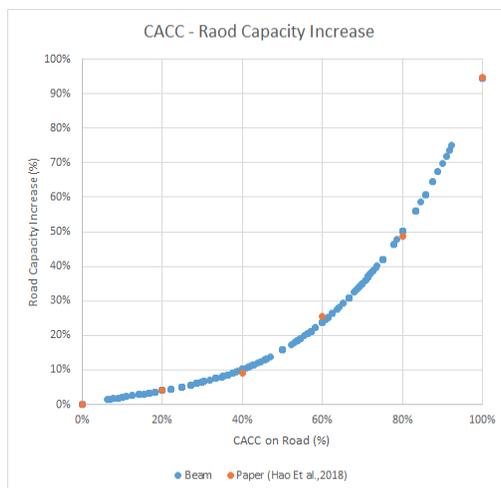

**Figure 2-2. Cooperative Adaptive Cruise Control impact on highway and arterial road capacity in BEAM is based on work by (Liu, Kan, Shladover, Lu, & Ferlis, 2018) who used microsimulation to estimate capacity increases as a function of the penetration of CACC-enabled vehicles on the roads**

## 2.5 Scoring and Replanning (Pre-day)

### 2.5.1 Clearing Modes and Routes

As is typical for MATSim simulations, after an iteration a portion of agents are randomly selected to modify their plans before the next iteration of AgentSim. In cases where agent mode choice is treated as fixed (for instance, in cases where activity plans and mode choices are imported as exogenous from a travel demand model), this replanning step means that some portion of agents will need to re-do route choice – with the rest of agents choosing a favorably performing plan from their library of past plans and taking the exact routes defined there. In cases where travel behavior is not treated as fixed, some agents are allowed to clear their chosen modes and/or discretionary activities (see Section 2.5.2) from their plans as well, re-making those choices in response to updated travel times and skims. Letting agents replan their route, mode, and/or activity choices allows the system to move towards user equilibrium, but limiting the number of agents who can replan in a given iteration reduces the risk of the system oscillating between non-optimal solutions.

When an agent's route is cleared, in the next iteration they make the same mode choice for each trip but, for any personal vehicle trips in their plans, they chose an updated lowest-utility (combining travel time and cost) route calculated given the current iteration's link speed table (which itself has been updated based on the traffic network simulation based on the path traversals from the previous iterations). This updating of routes, along with the ability to add random noise to the link speeds used for routing, is intended to spread travelers across many possible routes from an origin to a destination with similar travel times, rather than assigning all travelers between two points to the same optimal route.

After agents in personal vehicles reroute their trips, all agents with cleared modes are allowed to make a new mode choice for every trip, based on updated traffic network speeds for auto-based modes, as well as any additional information (such as parking availability, transit vehicle crowding, and ridehail wait time) that may have changed from the previous iteration. This updating of modes is also intended to avoid having the system oscillate between non-optimal solutions; for example, one where drive mode share is very low (and speeds are reported as fast for the next iteration) and one where drive mode share is high (and speeds are reported as slow for the next iteration).

### 2.5.2 Discretionary Activity Choice

BEAM also includes a rudimentary destination choice model that allows for agent activity plans, including the location, type, and timing of some activities, to change in response to system conditions. For the purpose of these simulations, BEAM divides activities into mandatory (work and school) activities, with fixed exogenous locations and start/end times, and discretionary (all other purposes, including leisure and shopping), which have the option of being updated endogenously. When the discretionary activity choice model is turned on, some agents are selected each iteration to clear all discretionary activities from their plans, keeping only mandatory activities. When the next iteration of AgentSim starts, these agents will be given the option of adding new discretionary activities to their chosen plan. For every mandatory activity, a single blank subtour (an outbound trip, an activity, and a return trip, with trip destination and timing left blank) is added to the plans. For agents without any mandatory trips, two more blank subtours are added (one in the morning, one mid-day, and one in the afternoon, returning home in between each), ensuring that every agent can make at least three discretionary subtours. The presence of these blank subtours in an agent's chosen plan signals that agent to begin their discretionary activity choice process during the next iteration.

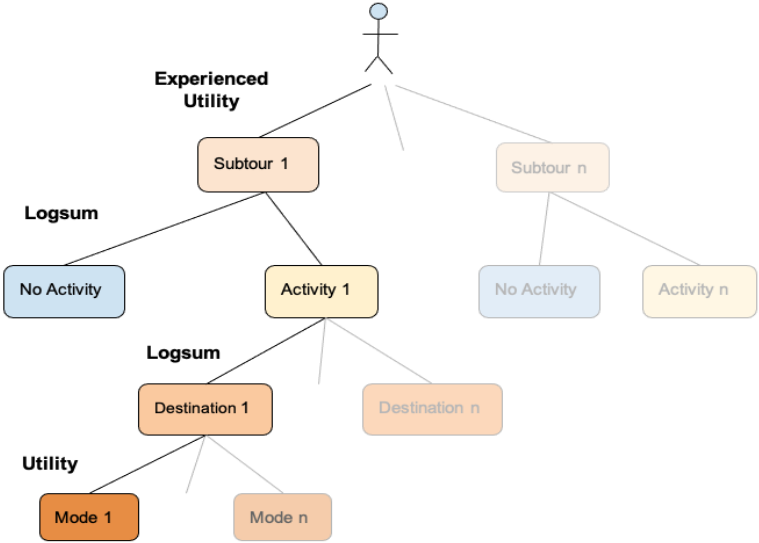

Figure 2-3. Diagram of the nested discretionary activity choice decision structure

When the person agents are loaded into AgentSim for the next iteration, details on any blank subtours are filled in. This discretionary activity choice process happens probabilistically based on the multimodal accessibility of potential trip destinations. The various choices involved in determining a discretionary subtour are modeled in a nesting structure, with the accessibility of each mode informing the favorability of each destination, and the favorability of each destination informing the decision of whether to take the trip at all. Allowing agents to replan their discretionary activities several times allows them to try different activity types and durations and be more likely to select successful plans. A schematic for this nested decision structure is shown in Figure 2-3.

Filling in details of subtours involves defining an activity type, activity timing and duration, and destination choice set for each blank discretionary subtour. The start time of the discretionary activity for each blank subtour is chosen from between the start and end time of the surrounding mandatory activity (padded by 30 minutes on either end), with the relative probability of each activity type/start hour combination defined by the input "activityIntercepts" file. The process of generating a subtour activity type and start time is defined in pseudocode below in Figure 2-4.

```
getDiscretionaryActivity(mandatoryStartTime, mandatoryEndTime, activityIntercepts):
    """
    Inputs:
        mandatoryStartTime -- start time (in hours) of the mandatory activity from
            which the discretionary subtour is being created
        mandatoryEndTime -- end time (in hours) of the mandatory activity from
            which the discretionary subtour is being created
        activityIntercepts -- m*n array where m is the number of discretionary
            activity types and n is the number of hour bins. The values of this
            array represent the relative likelihood of a discretionary activities
            of type associated with index m starting in time bin associated with
            index n. This is loaded from the activityIntecepts input file.

    Outputs:
        chosenActivityInd -- chosen activity type, determined by the index
            associated with the set of possible activities represented by index m
        chosenStartTimeInd -- hour of day of the discretionary activity start,
            determined by the index associated with the set of hours
            represented by index n
    """

    startInd = ceil(mandatoryStartTime + 0.5)
    endInd = floor(mandatoryEndTime - 0.5)
    chosenActivityInd, chosenStartTimeInd = randSample(activityIntercepts[:,
            startInd:endInd])
    return chosenActivityInd, chosenStartTimeInd + startInd
```

**Figure 2-4. Algorithm for generation of the subtour activity type and start time.**

Given the chosen time bin, the precise starting time (in seconds from midnight) is sampled uniformly from the chosen time bin. Given the chosen activity type, the activity duration is sampled from an exponential distribution with each activity's mean duration defined as an input in the activityParams input file.

The destination choice set for each discretionary tour is determined by sampling uniformly from all TAZs within a given distance of the tour origin, with both the number of potential destinations

TAZs and the maximum sampling distance defined in the scenario configuration file. This sampling is done with a fixed seed so that the discretionary activity location choice set for each potential discretionary tour is the same for a given agent from one iteration to the next[4].

At this point, the agent has chosen the activity type and activity start time for every blank subtour, and BEAM has generated a subsample of TAZs offering possible locations for this activity. The destination is chosen based on the expected maximum utility of traveling to and from that destination via each available mode. These utilities are estimated by looking up travel times and costs from BEAM's origin/destination skims (see Section 2.6.3) and then converting those values to a utility value based on the agent's utility functions; a penalty term is added if the return trip cannot be completed before the agent's next trip is scheduled to begin. The expected maximum utility for each potential tour is calculated slightly differently depending on whether tour mode choice is enabled in BEAM. If your mode choice is enabled, the expected maximum utility for a given tour destination is calculated as the logsum of the tour mode utilities as defined in Section 2.3.4. If tour mode choice is not enabled, the expected maximum utility of each destination is calculated as the $logsum$ of the utility associated with using each mode available to a household. In either case, the inclusive value of the $logsum$ of tour mode utilities (concretely, the benefit of having multiple good modal alternatives to a given destination) a configurable parameter. The expected maximum utility of making a subtour at all is calculated as the $logsum$ of these destination $logsums$, with a separate configurable parameter, plus the estimated utility of participating in the activity. This participation utility is defined as a constant plus an activity-specific parameter multiplied by the activity duration. This constant is defined in the scenario configuration file, and the activity-specific value of time parameters are defined in the activityParams input file.

The actual nested choices are then calculated for each subtour. The decision whether to take a trip at all is a binary logit comparing the expected maximum utility of taking a tour to the utility of not taking a tour. The utility of taking a tour is determined by adding a constant $β_{0,act}$ to the expected utility of participating in the activity (the mandatory activity's time utility coefficient $β_{t,act}$ multiplied by its duration $T_{act}$) plus the expected utility associated with traveling to and from the activity $\widetilde{U}_{act}$. By convention, the utility associated with not taking a trip is assumed to be zero.

$$U_{act} = β_{0,act} + T_{act}β_{t,act} + \widetilde{U}_{act}$$

$$U_{noact} = 0$$

The expected utility associated with travel $\widetilde{U}_{act}$ is taken for all modes over a sample of possible destinations and is calculated using a $logsum$:

---

[4] This random sampling of a destination choice set is done to mimic unobserved heterogeneity of the favorability of different destinations, which is assumed to be a fixed effect that is constant for every agent. If this effect were not fixed over iterations, the solution would eventually converge to destinations closer and closer to the tour origin.

$$\tilde{U}_{act} = \lambda_{dest} \log\log \left( \sum_{d \in dests} e^{\tilde{U}_d / \lambda_{dest}} \right)$$

where $\lambda_{dest}$ is the nest correlation and $\tilde{U}_d$ is the expected maximum utility of travel to and from destination d across all modes. This destination-specific utility is calculated as the $logsum$ of the utilities of travel to and from a given destination for each mode $m$, $U_{d,m}$:

$$\tilde{U}_d = \lambda_{mode} \log\log \left( \sum_{m \in modes} e^{U_{d,m} / \lambda_{mode}} \right)$$

$$U_{d,m} = \beta_m + \beta_{cost} C_{m,d} + \beta_{time,m} T_{m,d} + \beta_{transfer} Transfer_{m,d}$$

These mode and destination specific utilities are calculated using the same utility structure as the mode choice model, using expected travel times, costs, and transfers taken from the skims (see Section 2.6.3). If the alternative to take a tour is selected, the destination is chosen based on the utilities associated with the different destination alternatives. These choices–discretionary activity participation, location, and timing--are added to the agent's plans, and the simulation is then started[5]. The mode for each trip in the plans is determined on the fly by the typical ChoosesMode activity.

## 2.6 Model Outputs

When BEAM is run it produces disaggregate outputs sufficient to reproduce every action taken by each agent in the simulation, as well as several aggregated outputs that can be used to calculate relevant performance metrics and to serve as inputs for additional BEAM runs or for other models.

### 2.6.1 Events

The most disaggregated outputs returned by BEAM are in the form of an events file. BEAM is an event-based simulation where, rather than advancing in discrete time steps, the simulation allows agents to change state at any time, with a scheduler enforcing the sequencing of these state changes. Each of these state changes is recorded in the form of a simulation event, which can be collected and output at the end of a simulation. BEAM allows for many different types of events associated with different behaviors, adding attributes to the core set of MATSim events (such as path traversals and events associated with agents entering and exiting vehicles) as well as defining new ones, including those associated with parking and refueling. Table 2-3 summarizes the outputs included in the events file.

---

[5] Discretionary activity location choice and scheduling is done before an iteration, rather than on the fly during an iteration, to allow the household CAV scheduler to access and plan for discretionary trips.

**Table 2-3. Each simulation event in BEAM output is recorded with the time of the event and several fields specific to the event type.**

| Event Type | Selected Attributes | Description |
|---|---|---|
| **Path Traversal** | Duration, path, energy consumption, vehicle type, vehicle ID, mode | Produced each time a vehicle completes a route during AgentSim, e.g., at the end of the trip. Gives details on the vehicle, route, and passengers. |
| **Mode Choice** | Alternatives available, mode chosen, location, person ID | Produced each time a person departs for a trip, summarizing the modes available for the trip and the one chosen |
| **Replanning** | Replanning reason, initial mode, person ID | Gives the initial intended mode and the reason why the trip could not be completed |
| **Reserves Parking** | Parking TAZ, vehicle ID, person ID, parking cost, charger availability | Records information each time a vehicle enters a parking spot |
| **Charging Plug in/Out** | Vehicle ID, parking TAZ, charger type, initial/final state of charge | A record of each time a vehicle begins or ends refueling at a charging station |
| **Person Enters/Leaves Vehicle** | Vehicle type, vehicle ID, time, person ID | A record of every time an agent enters or exits a vehicle. Note: a walking trip is accomplished in a "body" vehicle that is entered as soon as an activity completes |
| **Activity Start/End** | Activity type, location, person ID | Records when agents begin and end activities |

These events can be analyzed to give aggregate system wide outcome measures. For instance, the distance field of path traversal events can be summed in order to give a measurement of the total vehicle miles traveled in the simulation, and if weighted by the number of passengers present it can instead give the total passenger miles traveled throughout the system. Outputs can also be aggregated across individual vehicles as they travel through specific TAZ and census tracts/block groups, in order to evaluate the spatial and temporal concentration of criteria pollutant emissions relative to the populations exposed to those emissions. These measures can be calculated separately for different modes, vehicle types, time periods, and demographic characteristics. Mode choice events can be aggregated to give the mode split across all trips or disaggregated by location, time of day, or trip distance.

### 2.6.2 Linkstats

The routable road network in BEAM is represented by a set of connected links with attributes governing their congestion characteristics (such as free flow speed and capacity, and optionally other parameters of the Bureau of Public Roads volume delay function) and the modes allowed on them. During an AgentSim iteration, the router uses a static representation of the travel speed in order to produce shortest path routes for potential trips on the network. This static representation of the routable street network allows for speeds to vary across a predefined set of time periods, defining a travel time for each link for each time period of the day. This representation is static in that the speeds within a time period remain fixed during an AgentSim iteration regardless of agents' decisions. Instead, link speeds are updated by PhysSim after an AgentSim iteration completes, and the timestep-averaged speeds produced by PhysSim for each link serve as the static input travel speeds for the next iteration of AgentSim.

This static representation of time-varying link travel times is output by PhysSim as a linkstats file. This linkstats file contains the congested travel time of each link for each time period used by the router for the next AgentSim iteration, as well as other information such as the link volume (the number of vehicles exiting the link during the time window) for both heavy-duty and light-duty vehicles. It is important to note that individual and aggregate travel times can differ between the AgentSim travel times reported in the events file and the PhysSim, as the AgentSim travel times reflect congestion patterns given the previous iteration's travel demand. In a simulation where a state approaching dynamic user equilibrium has been reached these differences will be small, but in a simulation that has not reached relaxation or has extreme pockets of unresolved congestion the differences between these measures can be substantial. Indeed, one useful measure for the extent to which a simulation has run for enough iterations is whether the gap between AgentSim and PhysSim speeds (or the difference between successive PhysSim speeds) has stabilized at a small value.

### 2.6.3 Skims

Some components of BEAM (such as pre-day activity planning module) and some linked models (such as ActivitySim) require a lightweight and fast way of accessing the expected travel time and other characteristics of a potential trip but do not require the level of detail produced by the router. These approximate trip characteristics are provided by the skims, a set of lookup tables indexed by trip mode, origin, destination, and time of day that estimate trip travel time, cost, and other relevant variables based on an average of those values for similar trips in the previous one or several iterations. Because the skimmer is structured as a lookup table, it can be accessed substantially more quickly than the time taken to calculate a new route, but because the geographic resolution is at that TAZ level, the results are less accurate. At the end of each AgentSim iteration, the travel time, cost, and other relevant characteristics of all trips sharing the same origin TAZ, destination TAZ, mode, and start time bin are averaged and saved into the skimmer for the next iteration. This aggregated lookup table can also be output at the end of an iteration as a long .csv file, allowing skims to be used as a "warm start" for a subsequent BEAM run or to inform a travel demand model such as ActivitySim.

BEAM also provides the ability to output trips that are indexed by a single (origin or destination) TAZ rather than an origin/destination pair. In particular, BEAM allows for the generation of separate parking skims and ridehail skims that further aggregate across all trips of a given mode in a given time period arriving to (departing from) a given TAZ regardless of trip origin (destination). The parking skims summarize average parking cost and inbound walk distance for a given destination TAZ at a given time of day, and the ridehail skims summarize average wait time, average per-mile cost, and portion of requests that cannot be successfully matched with a ridehail vehicle for a given origin TAZ and time of day.

### 2.6.4 Debugging plots and summary tables

To ease quick initial analysis of model performance, BEAM automatically produces a directory of figures and tables showing a rough picture of the simulation outcomes. These plots and tables tend not to be of high enough quality for publication directly, but give users an initial view into many of the high-level metrics that need to be verified to confirm that the model is operating as expected. These products include summaries of mode split, activity participation, parking and charging activity, energy use, ridehail service quality, road network speed, and model runtime. Figure 2-5 is an example of a figure produced by BEAM showing the number of completed trips by mode for each iteration of a simulation of 200,000 agents in the New York City metropolitan area. Note that the total number of trips taken sharply drops and then gradually increases after the first iteration as agents naively attempt many discretionary tours, experience poor results, and then gradually learn more favorable activity patterns. Table 2-4 is an example table produced by BEAM for the same New York run, showing several selected performance metrics and how they change from iteration to iteration.

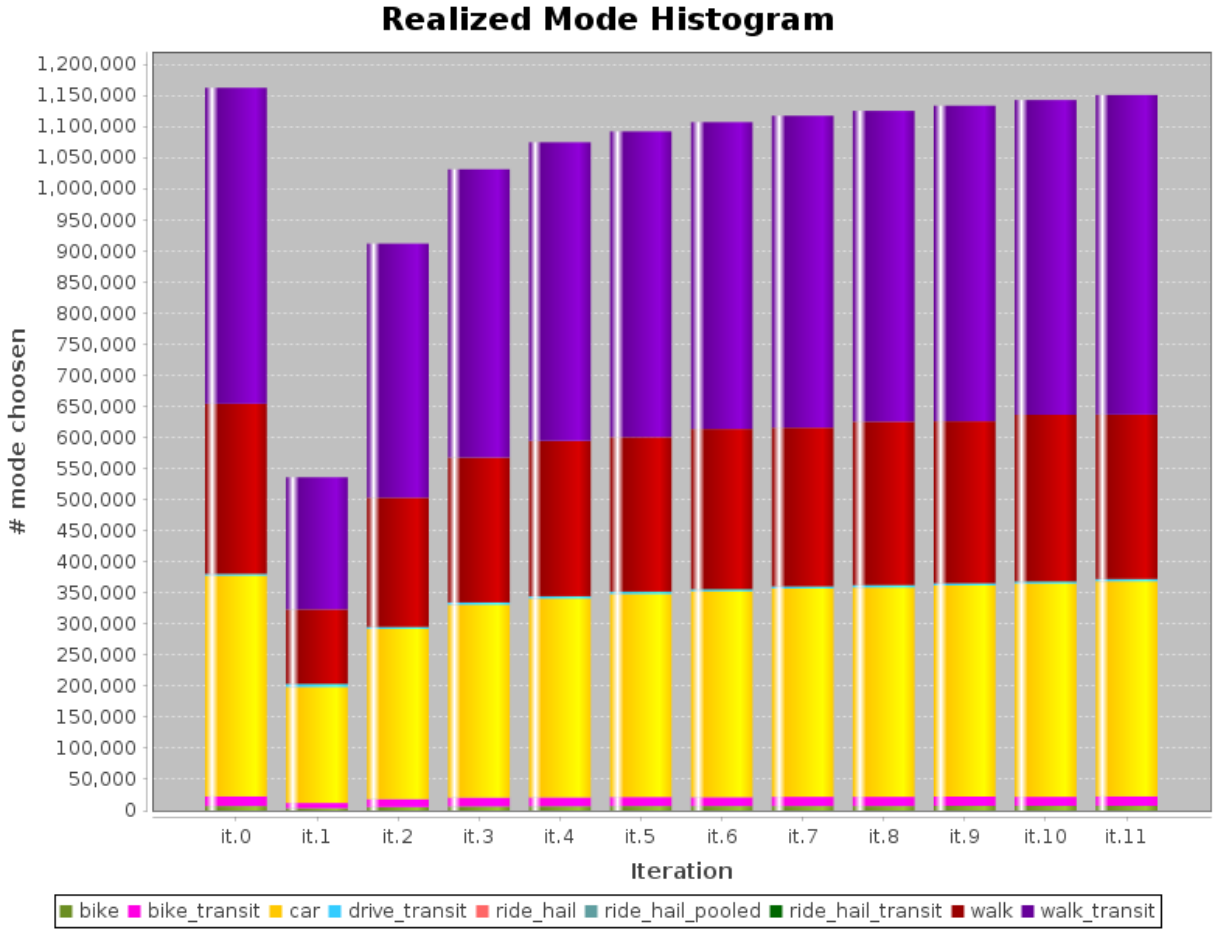

**Figure 2-5. Sample figure produced by BEAM showing the number of completed trips by mode for each iteration (simulation of 200,000 agents in the New York City metropolitan area).**

**Table 2-4. Sample table produced by BEAM for the same New York run.**

| Iteration | Unmatched Ridehail Trips | Average Ridehail Wait (min) | Gasoline consumption (GJ) | Total Vehicle Delay (hrs) | Agent Time On Crowded Transit (hrs) |
|---|---|---|---|---|---|
| 0 | 0.34% | 4.5 | 12,823 | 44,292 | 396,563 |
| 1 | 1.28% | 6.1 | 9,578 | 45,426 | 184,859 |

| Iteration | Unmatched Ridehail Trips | Average Ridehail Wait (min) | Gasoline consumption (GJ) | Total Vehicle Delay (hrs) | Agent Time On Crowded Transit (hrs) |
|---|---|---|---|---|---|
| 2 | 0.83% | 4.9 | 11,037 | 45,104 | 288,737 |
| 3 | 0.53% | 6.6 | 11,896 | 45,617 | 312,329 |
| 4 | 0.18% | 7.3 | 12,064 | 45,484 | 327,168 |
| 5 | 0.58% | 7.3 | 12,226 | 45,534 | 325,639 |
| 6 | 0.60% | 6.2 | 12,299 | 45,504 | 333,308 |
| 7 | 0.38% | 6.7 | 12,411 | 45,541 | 331,624 |
| 8 | 0.77% | 5.7 | 12,458 | 45,518 | 338,534 |
| 9 | 0.56% | 7.4 | 12,510 | 45,525 | 335,479 |
| 10 | 0.20% | 7.4 | 12,600 | 45,535 | 344,378 |

# 3. Model Linkages

## 3.1 Population Synthesis

BEAM has been developed to read in population files generated by SynthPop (Ye, Konduri, Pendyala, Sana, & Waddell, 2009), an open-source Python implementation of PopGen that generates a synthetic population for a given region. SynthPop relies on data from the U.S. American Community Survey (ACS) and Census Public Use Microdata Sample (PUMS) that can be directly downloaded using U.S. Census APIs. SynthPop defines a synthetic population of individuals, including demographic characteristics, and assigns them to households, which can be given attributes such as number of vehicles owned and household income. The final set of sociodemographic variables used can be chosen from those available in ACS and PUMS to fit those needed by the behavioral models used in BEAM and elsewhere.

## 3.2 Pre-day Planning

In order to run its internal discretionary activity choice models, BEAM relies on agents already having mandatory trips defined. This can be accomplished for a Synthpop population using a lightweight tool developed at UC Berkeley called ActivitySynth (UrbanSim, 2023), which implements a model of workplace location and departure time choice estimated on California Household Survey data. This tool outputs plans for all workers in the synthetic population that give the location and start/end times of the primary work activity. Mandatory trips serve as the starting point for BEAM's internal activity modeling described in the section Discretionary Activity Choice.

If a fully featured activity-based model is required, BEAM can assign agents' plans developed by the ActivitySim model. ActivitySim allows for the generation of plans using rich behavioral logic, including joint tours, mandatory trips to schools, and intermediate stops. An implementation in the San Francisco Bay Area that has been modified to directly produce output plans in a format readable by BEAM has been developed at UC Berkeley (Galli, et al., 2009). When BEAM loads ActivitySim plans, it can be configured either: to keep ActivitySim's mode choices for all simulated trips (thus only simulating downstream choices, such as route, parking, and charging choices); to keep each agent's planned activities but rely on BEAM's internal models for mode choice; or to start with ActivitySim's discretionary activity and mode choices and subsequently update both during the replanning phase. The relative benefits of each method depend on the degree of behavioral sensitivity a user intends to simulate. For studies focusing on charging behavior or the operations of small on-demand fleets, for instance, it is likely appropriate to treat activity and mode choices coming from ActivitySim as fixed. For studies where small but systematic changes to mode choice are expected (for instance, increasing the size or changing the price of an on-demand fleet or transit service), it is likely appropriate to allow BEAM to update agent mode choices while retaining discretionary activity choices. For a situation involving substantial changes to travel behavior (such as introduction of an entirely new travel mode or the onset of a global pandemic), or in a situation where outputs from an activity-based travel demand model are not available at all, it likely makes sense to use BEAM's discretionary activity model.

## 3.3 Energy Consumption

Vehicles in BEAM are initialized with a primary fuel type, energy consumption per unit distance, and fuel capacity. Vehicles can be assumed to begin an iteration fully fueled, or their starting fuel level can be drawn from a uniform distribution with configurable parameters. Optionally, a vehicle can also be given a secondary fuel type, energy consumption rate, and fuel capacity. In these cases, such as for plug-in hybrid electric powertrains, vehicles are assumed to consume their primary fuel type when there is any available, and then use the secondary fuel type after the primary fuel type has been depleted. In the simulation, a vehicle's current state of charge or fuel level influences parking location choice, the timing of any required refueling, and the amount of fuel put into the vehicle during refueling.

## 3.4 Performance Metrics

Because it produces link-based speeds as well as parking costs and wait times and costs for on-demand modes, BEAM produces outputs that are sufficient for the calculation of the Mobility Energy Productivity (MEP) metric, a measure of the proximity of a given location, and energy required to reach, a collection of activities (Hou, Garikapati, Nag, Young, & Grushka, 2019); as well as the Individual Experienced Utility-based Synthesis (INEXUS), which is complementary to, but distinct from, the location-based MEP, and calculates the travel utility of each individual agent under baseline conditions and different scenarios (Garikapati, 2023).

## 3.5 Power Grid Co-Simulation

Charging battery electric vehicles in BEAM is centralized in the Actor ChargingNetworkManager. The latter implements HELICS (Panossian, et al., 2023) a co-simulation framework that allows BEAM to run simultaneously with other simulators and exchange states in real time. Under this structure, BEAM is able to interact with a power grid model, by communicating observed loads and expected physical bounds for load management.

# 4. Model Deployment

For any simulation model it is important to provide pipelines to increase the efficiency of model scenario setup and deployment. The BEAM code base provides a variety of tools to prepare the scenario data (network, population, households, vehicles, etc.). Typically, models are run with a baseline model and alternative scenarios in mind. The differentiation between the models is either in the value of parameters e.g., road capacities, supplies of ride-hail vehicles, etc. or is some new feature which has been implemented specifically for the scenario under consideration. Due to the size and runtime of a single simulation, it is often not possible to run larger simulations locally on laptops or desktops. Therefore, multiple tools and pipelines have been developed which allow BEAM simulations to be deployed either at a high-performance computing cluster such as on the National Energy Research Scientific Computing Center (NERSC)[6] or on commercial cloud computing services such as Amazon Web Services (AWS) or Google Cloud Engine (GCE). With a single command, the simulation can be deployed using the specified configuration file. Based on scenario size, memory or computational resources can be specified and which code to execute with what data from which repository and version. Once a simulation is started, the simulation output can be accessed and analyzed using tools such as JupyterLab[7], which allows quick adaptations to the simulation execution plan. Furthermore, various data and visualizations are automatically generated from the simulation output files, facilitating the monitoring of additional simulations. After a simulation is completed, the outputs are automatically archived before the simulation machine is shut down.

## 5. Case Study: New York City

This section summarizes the steps taken to calibrate and validate the baseline BEAM model for the New York City region in January 2020 (pre- COVID pandemic). This is an example where a BEAM implementation was developed from scratch, without an existing activity-based travel demand model to provide plans and using only minimal employment data to generate agents' mandatory plans. It is a proof-of-concept that BEAM can simulate consistent, realistic transportation system outcomes and serve as a testbed for simulating major mobility shifts in a way that is sensitive to congestion, multimodal accessibility, and changes in activity patterns. While these results have been validated against high level metrics, they are not intended to replace detailed travel demand models of the sort developed by Metropolitan Planning Organizations that take years of data collection, calibration, and validation. The application of this model to COVID recovery scenarios, which required capturing all of the endogenous behavior change described above, will be described in a forthcoming report. This report presents the baseline scenario to illustrate BEAM's capabilities in practice. First the simulated baseline is described, followed by a detailed description of the calibration and validation of the baseline.

### 5.1 Scenario Definition

---

[6] The National Energy Research Scientific Computing Center (NERSC) is the primary scientific computing facility for the Office of Science in the U.S. Department of Energy (https://www.nersc.gov/)

[7] JupyterLab is the latest web-based interactive development environment for notebooks, code, and data. Its flexible interface allows users to configure and arrange workflows in data science, scientific computing, computational journalism, and machine learning (https://jupyter.org/)

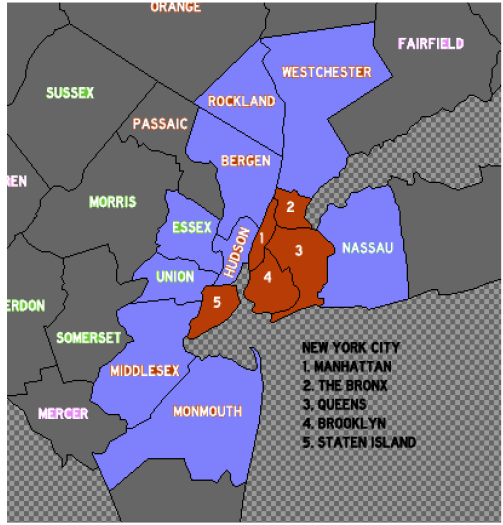

**Figure 5-1. BEAM modeling domain: Five boroughs of New York City and nine counties as the study area.**

The baseline is intended to simulate the travel during an average weekday in the New York City region; the five boroughs plus nine outlying counties (Essex, Union, Hudson, Middlesex, Monmouth, Bergen, Rockland, Westchester, and Nassau). The covered area (see Figure 5-1), includes a total population of about 13 million inhabitants. The travel of all agents is simulated across 3890 Traffic Assignment Zones (TAZs) defined for the modeling domain as seen in the left map of Figure 5-2. Parking availability was estimated using publicly available datasets on street parking and off-street garages, and the road network was taken from OpenStreetMap and simplified using the OSMnx tool.

The baseline contains a total of about 36.5 million trips, with an average of about 4 trips per day per agent, and a total of about 44 million vehicle miles traveled, including both personal, transit, and shared vehicles, consuming about 84 terajoules of energy, and a total of about 73 million personal miles traveled. The relative intensity of the demand is depicted by the heatmap of the most visited locations in Figure 5-2.

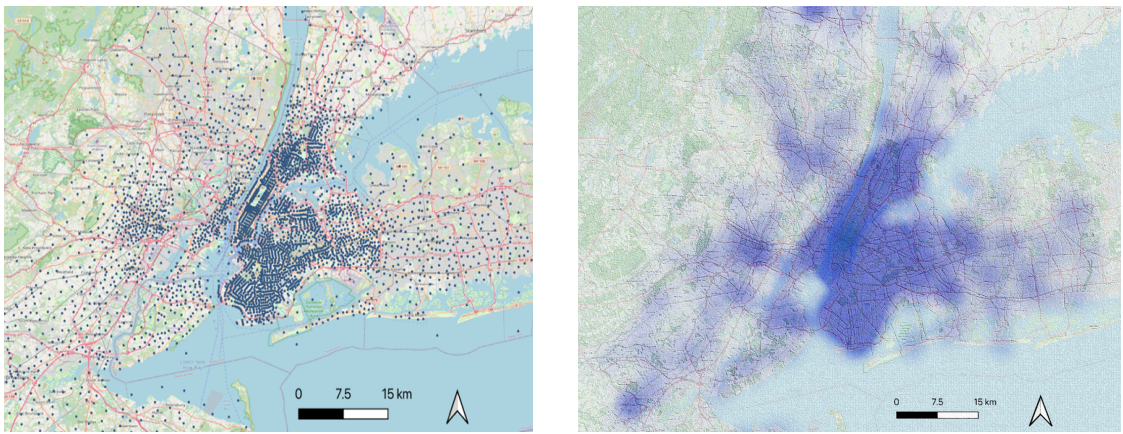

**Figure 5-2. BEAM modeling domain (Left) Centroids of the considered TAZs; (Right) Heatmap showing the more visited locations.**

The initial population was generated using the SynthPop tool, based on Public Use Micro (PUMS) data from the U.S. Census for the New York metro area, with its default configurations modified to include information about workers' work industry in the population outputs and to include aggregate information on workers by job category in the block group marginals being targeted. Given each worker's home location and work industry, workplace locations are determined by sampling from the American Community Survey commuting flows. Work departure and arrival times are sampled from the NHTS, with the distribution taken across the study area.

## 5.2 Calibration

Once the population and activity plans were created, to make the process more time- and cost-efficient, the New York City scenario calibration was performed on a smaller (4%) sample of the active travelers, and then applied to a larger sample (10%), with the capacity of each transit vehicle reduced to 13% of it's rated capacity (with the extra 3% added to account for sampling variability).

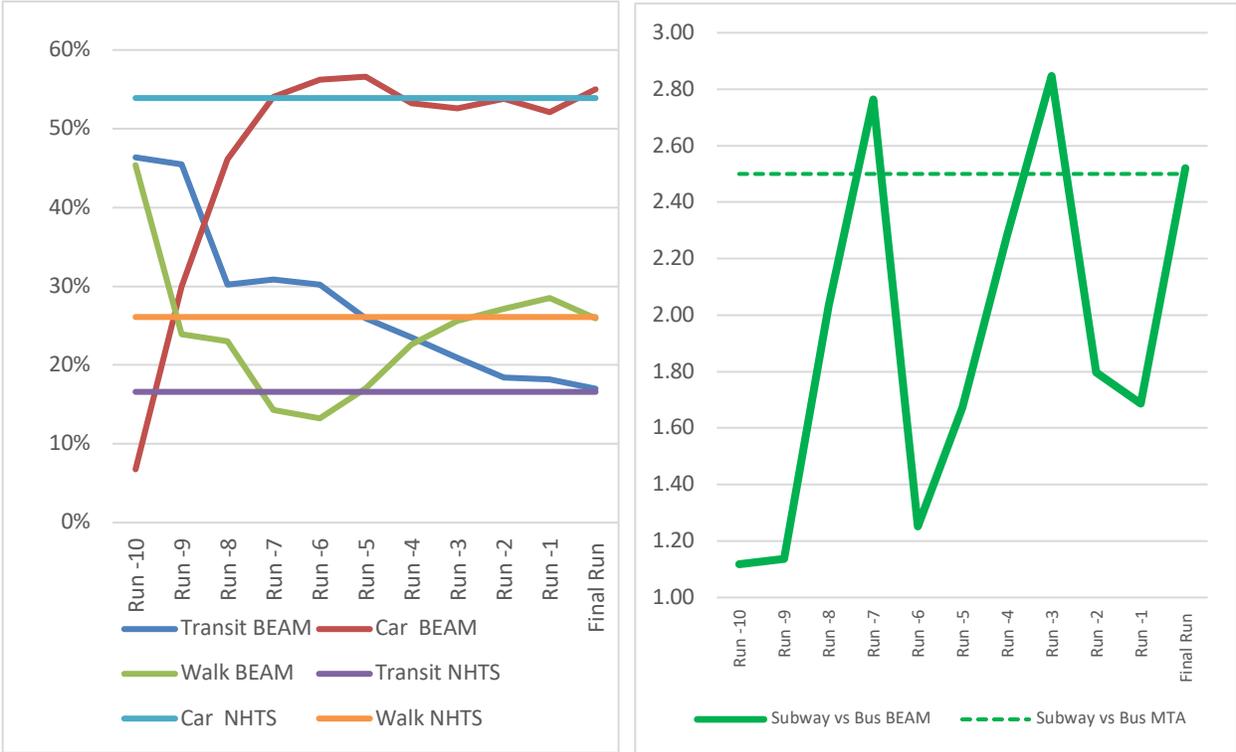

**Figure 5-3. (Left) Mode Split: BEAM simulated data (scaled 4% samples) vs NHTS observed data; (Right) Subway vs Bus ratio for the MTA agency**

The calibration started using configuration parameters similar to other calibrated scenarios, like the San Francisco Bay Area scenario, and some of those have been varied to match specific local observed data. In particular, for this scenario, the calibration process mainly focused on the mode choice model and aimed at matching the mode split for the New York City CBSA in the 2017 National Household Travel Survey (NHTS), as well as the ratio of 2019 average weekday subway vs bus ridership provided by the New York Metropolitan Transportation Authority (NYMTA). In addition, the process to generate discretionary activities in BEAM was optimized to reproduce the average number of trips per person for specific purposes, again provided by NHTS. The parameters used to match the simulated data with the observed data were mainly the intercepts and value of time of the multinomial logit model used for the mode-choice, and the discretionary activity generation parameters.

The calibration process consisted of an iteration of four main steps: 1) Running the simulation: 2) Calculating the values to be compared with the observed data; 3) Comparing the simulated data with the observed data; 4) Based on the comparison results, adjusting the parameters to be used for the next iteration. Figure 5-3 shows how the simulated values were approaching the observed data every new iteration of the calibration process; in the figure the simulated data have been linearly scaled from the 4% sample to represent the full population. The mode split and ratio of subway to bus ridership were fully calibrated after 11 calibration runs. Note that while the simulated modal split gradually approached the observed data, the subway versus bus ratio was close to observed even during intermediate calibrations, showing how a transportation system is complex and output values are correlated: changing certain parameters during the calibration process can cause collateral effects on other variables.

**Table 5-1. Result of the calibration: simulated (BEAM) vs observed (NHTS) modal and activity splits for the full sample**

| | Mode Share | | | Trips per Person | |
|---|---|---|---|---|---|
| **Travel Mode** | BEAM | NHTS | **Activity Type** | BEAM | NHTS |
| **Transit** | 18.5% | 16.6% | Home | 1.73 | 1.88 |
| **Bike** | 0.8% | 1.0% | Meal | 0.28 | 0.32 |
| **Car** | 50.8% | 53.9% | Shopping | 0.29 | 0.33 |
| **Taxi** | 0.8% | 1.8% | Social/Recreation | 0.41 | 0.43 |
| **Walk** | 29.1% | 26.1% | Work | 1.03 | 0.99 |
| **Other** | 0.0% | 0.6% | Other | 0.33 | 0.39 |

Table 5-1 shows the results of the calibration process in terms of simulated versus observed values of the modal split and activity split for the full population, after the ten subsamples have been simulated with the calibrated parameters and their results merged together, confirming the transferability of the results from the ten subsamples to the full population. Regarding the ratio of subway to bus ridership for NYMTA, the final calibration value of 2.5 is exactly as observed.

## 5.3 Validation

Once the scenario baseline was calibrated, other observed data were used to validate the baseline: 2019 average weekday transit ridership on bus, subway, and commuter rail routes provided by NYMTA, Port Authority of New York and New Jersey (PATH), and New Jersey Transit (NJ Transit), as well as the overall modal split for different ranges of trip distance. The data used for validating the baseline are not correlated with the data used for calibration because even if the overall modal split is correct, it does not necessarily follow that the modal split is also correct across different trip distances. And if the transit share and transit vehicle splits from the calibration exercise are correct, it does not necessarily follow that the absolute ridership for each agency matches observed ridership.

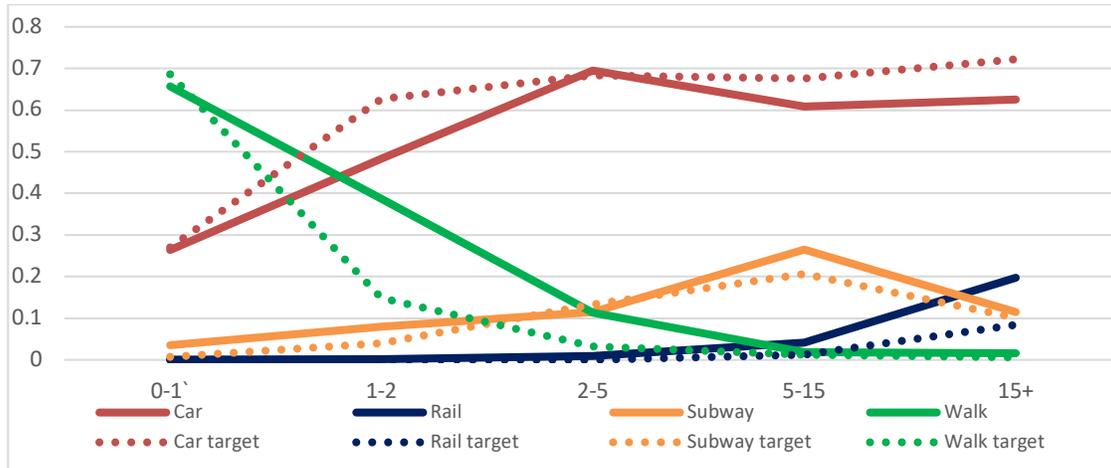

**Figure 5-4. Mode share as a function of trip distance: simulated versus observed data**

Figure 5-4 shows the modal split for each distance range for the travel modes with the most trips, confirming that the simulated trips correspond quite well to the observed data, thus indicating that the mode share model is well calibrated and correctly considers the travel distance cost for different modes of travel.

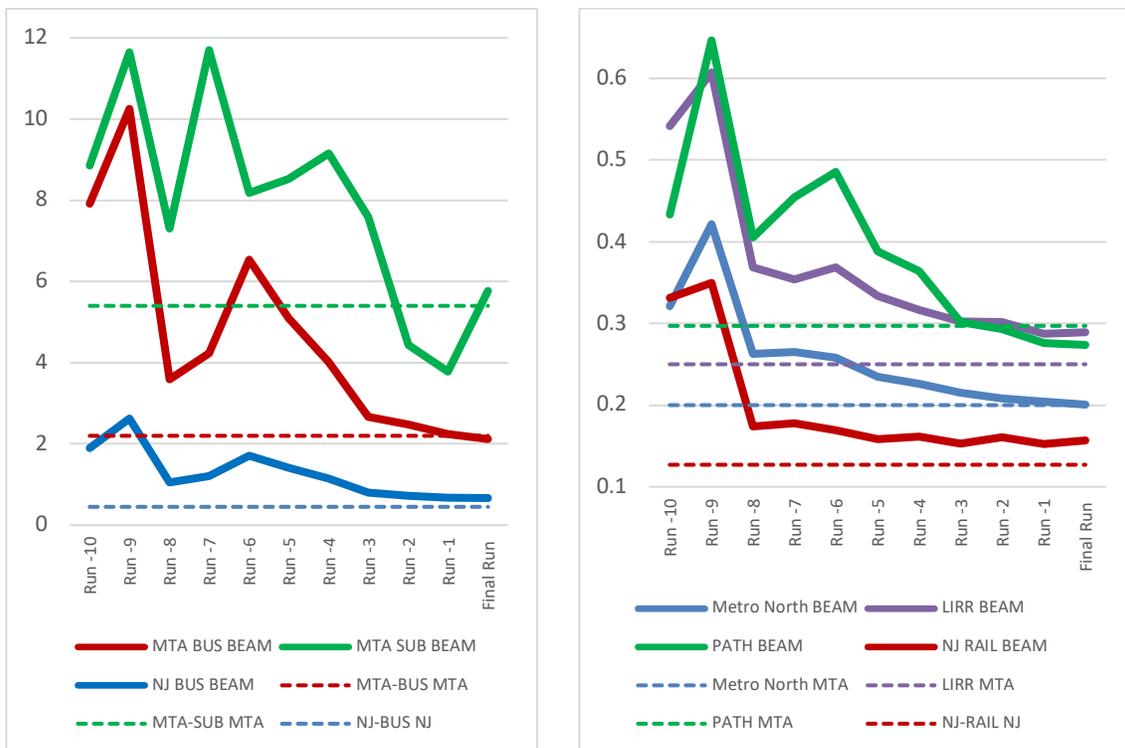

**Figure 5-5. Ridership in million passengers: BEAM simulated data (scaled 4% samples) vs 2019 average daily ridership from NYMTA, PATH and NJ Transit.**

BEAM: The Modeling Framework for Behavior, Energy, Autonomy & Mobility │33

Figure 5-5 shows that the simulated transit ridership of the main agencies, after scaling to the 100% population, changed with each calibration iteration until the observed NHTS value was achieved, thus confirming the validity of the baseline based on the observed data. Note that the transit ridership values used to validate the scenario are absolute ridership, thus confirming that the population sample and the created activity and travel plans are of a comparable magnitude.

**Table 5-2. Result of the validation: simulated (BEAM) vs observed (NYMTA, PATH and NJ) 2019 average daily ridership for the full sample**

| 2019 Average Daily Ridership | Simulated in BEAM | Observed |
| --- | --- | --- |
| MTA bus | 2.1M | 2.1M |
| MTA subway | 5.2M | 5.4M |
| Ratio of Subway to Bus rides | 2.5 | 2.5 |
| MTA Metro North | 207k | 200k |
| MTA LIRR | 302k | 250k |
| PATH commuter rail | 270k | 297k |
| NJ Transit bus | 692k | 451k |
| NJ Transit commuter rail | 177k | 127k |
| MTA bridges/tunnels | 346k | 302k |

Finally, Table 5-2 shows the fully calibrated simulated ridership data from the full population of agents, compared with the observed ridership data, confirming the validity of the full sample.